\documentstyle[epsfig]{ar}

\def\Journal#1#2#3#4{{#1} {\bf #2}, {#3} (#4)}

\def\NIMA{{\em Nucl. Instrum. Methods} A}
\def\NPB{{\em Nucl. Phys.} B}
\def\NPA{{\em Nucl. Phys.} A}
\def\PL{{\em Phys. Lett.}}
\def\PLB{{\em Phys. Lett.}  B}
\def\PRL{\em Phys. Rev. Lett.}
\def\PRD{{\em Phys. Rev.} D}
\def\ZPC{{\em Z. Phys.} C}
\def\EPC{{\em Eur. Phys. J.} C}
\def\JPG{{\em J. Phys.} G}

\def\ra{\rightarrow}
\def\be{\begin{equation}}
\def\ee{\end{equation}}
\def\bea{\begin{eqnarray}}
\def\eea{\end{eqnarray}}
\def\mrm{\mathrm}
\def\ol{\overline}
\def\etal{\mbox{{\it et al.}}}
\def\ie{\mbox{{\it i.e.}}}
\def\eg{\mbox{{\it e.g.}}}
\def\ibid{\mbox{{\it ibid}}}

\newcommand{\inmath}[1] {\ifmmode#1\else$#1$\fi}
\newcommand{\definmath}[2] {\def#1{\ifmmode#2\else$#2$\fi}}

\newcommand{\epem}{\mbox{$e^+e^-$}}
\newcommand{\mpmm}{\mbox{$\mu^+\mu^-$}}

\newcommand{\lplm}{\mbox{$\ell^+\ell^-$}}
\newcommand{\Zz}{\mbox{$Z^0$}}
\newcommand{\WW}{\mbox{$W^+W^-$}}
\newcommand{\Zg}{\mbox{$Z/\gamma$}}
\newcommand{\eeWW}{\mbox{$\epem\rightarrow\WW$}}
\newcommand{\qq}{\mbox{$q\overline{q}$}}
\newcommand{\ff}{\mbox{$f\overline{f}$}}
\newcommand{\lnu}{\mbox{$\ell\overline{\nu}$}}
\newcommand{\enu}{\mbox{$e\overline{\nu}$}}
\newcommand{\mnu}{\mbox{$\mu\overline{\nu}$}}
\newcommand{\tnu}{\mbox{$\tau\overline{\nu}$}}
\newcommand{\WWqqqq}{\mbox{\WW$\rightarrow$\qq\qq}}
\newcommand{\WWqqln}{\mbox{\WW$\rightarrow$\qq\lnu}}
\newcommand{\WWqqen}{\mbox{\WW$\rightarrow$\qq\enu}}
\newcommand{\WWqqmn}{\mbox{\WW$\rightarrow$\qq\mnu}}
\newcommand{\WWqqtn}{\mbox{\WW$\rightarrow$\qq\tnu}}
\newcommand{\WWlnln}{\mbox{\WW$\rightarrow$\lnu\lnu}}
\newcommand{\Wenu}{\mbox{$W\enu$}}
\newcommand{\Zzee}{\mbox{$\Zz\epem$}}
\newcommand{\Zqq}{\mbox{$\Zg\rightarrow\qq$}}
\newcommand{\xseeww}{\mbox{$\sigma\left(\eeWW\right)$}}
\newcommand{\xsww}{\mbox{$\sigma_{WW}$}}

\newcommand{\Mz}{\mbox{$M_Z$}}
\newcommand{\Mw}{\mbox{$M_W$}}
\newcommand{\Mh}{\mbox{$M_H$}}
\newcommand{\Gw}{\mbox{$\Gamma_W$}}
\newcommand{\Ebm}{\mbox{$E_{bm}$}}
\newcommand{\roots}{\mbox{$\sqrt{s}$}}

\newcommand{\mrec}{\mbox{$m_{rec}$}}
\newcommand{\mreca}{\mbox{$m_{rec_1}$}}
\newcommand{\mrecb}{\mbox{$m_{rec_2}$}}
\newcommand{\com}{center-of-mass}

\newcommand{\pT}{p_T}
\newcommand{\vpT}{\vec \pT}
\newcommand{\vp}{\vec p}
\newcommand{\uT}{u_T}
\newcommand{\vuT}{\vec\uT}

\newcommand{\mT}{m_T}

\newcommand{\pp}{p\overline{p}}
\newcommand{\SppS}{$S\pp S$}
\newcommand{\Zll}{Z\to\ell^+\ell^-}
\newcommand{\Zee}{Z\to e^+e^-}
\newcommand{\Zuu}{Z\to\mu^+\mu^-}
\newcommand{\ev}{e\overline{\nu}}
\newcommand{\uv}{\mu\overline{\nu}}
\newcommand{\lv}{\ell\overline{\nu}}
\newcommand{\Wlv}{W\to\lv}
\newcommand{\Wev}{W\to\ev}
\newcommand{\Wuv}{W\to\uv}

\newcommand{\Mt}{\mbox{$m_{top}$}}

\newcommand{\Opal}{\mbox{O{\sc pal}}}
\newcommand{\Aleph}{\mbox{A{\sc leph}}}
\newcommand{\Delphi}{\mbox{D{\sc elphi}}}
\newcommand{\Lt}{\mbox{L3}}
\newcommand{\CDF}{\mbox{CDF}}
\newcommand{\Dz}{\mbox{D\O}}
\newcommand{\UAO}{\mbox{UA1}}
\newcommand{\UAT}{\mbox{UA2}}
\newcommand{\SLD}{\mbox{SLD}}
\newcommand{\cdf}{\CDF\ Collaboration}
\newcommand{\uaone}{\UAO\ Collaboration}
\newcommand{\uatwo}{\UAT\ Collaboration}
\newcommand{\dzero}{\Dz\ Collaboration}

\newcommand{\Pythia}{\mbox{P{\sc ythia}}}
\newcommand{\Herwig}{\mbox{H{\sc erwig}}}
\newcommand{\Ariadne}{\mbox{A{\sc riadne}}}
\newcommand{\Excalibur}{\mbox{E{\sc xcalibur}}}
\newcommand{\Koralw}{\mbox{K{\sc oralw}}}
\newcommand{\Gentle}{\mbox{G{\sc entle}}}

\begin{document}
\renewcommand\textfraction{0.5}

\title{Precision Measurements of the $W$-Boson Mass}
\author{
 Douglas A.\ Glenzinski
   \affiliation{Enrico Fermi Institute, University of Chicago, Chicago, IL 60637 and Fermi National Accelerator Laboratory, Batavia, IL 60510}
 Ulrich Heintz
   \affiliation{Physics Department, Boston University, Boston, MA 02215}}
\markboth{Glenzinski \& Heintz}{Precision Measurements of the $W$-Boson Mass}


\begin{abstract}
The Standard Model of electroweak interactions has had great success in
describing the observed data over the last three decades.  The precision of
experimental measurements affords tests of the Standard Model at the quantum 
loop level beyond leading order. Despite this great success it is important 
to continue confronting experimental measurements with the Standard Model 
predictions as any deviation would signal new physics.  As a fundamental 
parameter of the Standard Model, the mass of the $W$-boson, \Mw, is of 
particular importance.  Aside from being an important test of the SM itself, 
a precision measurement of \Mw\ can be used to constrain the 
mass of the Higgs boson, \Mh.
In this article we review the
principal experimental techniques for determining \Mw\ and discuss their 
combination into a single precision \Mw\ measurement, which is then used to 
yield constraints on \Mh.  We conclude by briefly discussing future prospects
for precision measurements of the $W$-boson mass.
\end{abstract}

\maketitle


\section{INTRODUCTION}
\label{sec:intro}

The Standard Model of electroweak interactions (SM) theoretically unites the
electromagnetic and weak forces of nature.  It 
postulates that these forces are communicated between the constituent particles
of nature, quarks and leptons, by carriers known as gauge bosons.  In
particular, the electromagnetic force is carried by the photon, $\gamma$, while
the weak force is mediated by the neutral $Z$-boson, $Z^0$, and the charged
$W$-bosons, $W^\pm$.  As such, the $W$-boson is fundamental to the Standard 
Model.  Moreover, the mass of the $W$-boson, \Mw, is a parameter of the
theory itself, so that a comparison between the experimentally determined \Mw\
and the SM prediction provides an important and fundamental test of the  
theory.  Alternatively, a precision measurement of \Mw\ can be used to 
estimate, within the framework of the SM, other parameters such as the mass
of the Higgs boson, \Mh.

\subsection{Historical Overview}

The weak force was first inferred from observations of nuclear $\beta$-decay, 
$n \ra p + e^- +\ol{\nu}_e-$.  In 1935, Fermi postulated the first theory of
weak interactions.  The form of the interaction was taken to be analogous with
that of the electromagnetic interaction, and was characterized by a 
``coupling'' (or strength) parameter --- the Fermi constant, $G_F$.  
By comparing interaction rates, the strength of the weak force was estimated to
be about $10^{-5}$ that of the electromagnetic force.  Fermi's theory very
successfully described low energy weak interactions, but violated unitarity
at high energy. 

In 1967 Glashow, Weinberg and Salam proposed the electroweak $SU(2)\times U(1)$
gauge theory, which unifies the weak and electromagnetic forces~\cite{ewk}.  
The theory postulated that the weak force was mediated by massive particles,
the $W$- and $Z$-bosons, and predicted their masses to be of order $10^2$~GeV 
\footnote{Here and throughout this article we use units of $\hbar=c=1$}.  The 
discovery of the $W$-boson in 1983, with a mass of $81\pm5$~GeV 
\cite{Wdiscovery}, was a great success for the electroweak theory.  More 
rigorous tests of the theory required more precise determinations of the boson
masses.

Over the past 15 years a variety of experiments have measured the mass of the 
$W$-boson with ever improving precision.  The first measurements were made at
the CERN $S\pp S$ collider~\cite{SppS} by the \UAO~\cite{UA1} and 
\UAT~\cite{UA2} experiments.
The \UAT\ experiment made the first measurement of the $W$-boson mass at a 
relative precision below $1\%$~\cite{UA2-90}.  The \CDF~\cite{CDF} and 
\Dz~\cite{D0} experiments at the Fermilab TeVatron~\cite{TeV}, another 
$\pp$ collider, were the first 
to push the precision to the $0.1\%$ level.  More recently, measurements made
at the CERN $\epem$-collider, LEP, by the \Aleph~\cite{aleph}, 
\Delphi~\cite{delphi}, \Lt~\cite{lthree}, and \Opal~\cite{opal} experiments,
have also reached relative precisions of $0.1\%$.  Combining all these 
measurements yields a relative precision of $0.05\%$ and affords stringent
tests of the Standard Model.  In particular, due to radiative corrections,
such precision measurements offer indirect constraints on the mass of the
Higgs boson.

\subsection{The Electroweak Theory}

In the $SU(2)\times U(1)$ electroweak theory,  local gauge invariance is 
achieved by introducing four massless bosons, an isovector triplet
${\mathbf{W}}^\mu=(W^\mu_0,W^\mu_1,W^\mu_2)$, and an isosinglet, $B^\mu_0$. 
In analogy to the electromagnetic case, the electroweak Lagrangian can be 
expressed as a product of currents and coupling parameters:
\begin{equation}
  {\mathcal{L}} = g {\mathbf{J}}^{\mu}\cdot{\mathbf{W}}^{\mu} 
              + g^{\prime}J^{\mu}_{Y}B^\mu_0
  \label{eq:ewklagrangian}
\end{equation}
where ${\mathbf{J}^\mu}$ and $J^{\mu}_{Y}$ are the weak isospin and hypercharge
currents of the physical fermions (\ie\ quarks and leptons), respectively, and
$g$ and $g^\prime$ are their couplings to the $\mathbf{W}^\mu$ and $B^\mu_0$ 
fields.  The weak quantum numbers are related to the electric charge, $Q$, by
$Q = I_3 + Y/2$, where $I_3$ is the third component of the weak isospin 
associated with the $SU(2)$ group and $Y$ is the weak hypercharge associated 
with the $U(1)$ group~\cite{ewktexts}.  The fact that the associated bosons 
are massless implies that the weak field is a long-range (infinite) field, in 
contradiction with experimental evidence.  This short-coming can be addressed 
by imparting mass to the vector bosons, which is achieved by spontaneously 
breaking the $SU(2)\times U(1)$ symmetry with the introduction of an additional
field.  Demanding that the theory be valid to high energies and remain 
renormalizable, a necessary condition in order to extract meaningful 
theoretical predictions, constrains the form of this additional field.  The 
simplest solution introduces a complex scalar isodoublet, the Higgs field, with
one component having a vacuum expectation value $v > 0$~\cite{higgs}.  The 
physical boson fields can then be expressed as
\begin{equation}
  W^\pm = (W_1 \pm W_2)/\sqrt{2},\:\:\:
  \left(\begin{array}{c} Z^0 \\ A^0 \end{array} \right) =
  \left( \begin{array}{cc}
           \cos\theta_W & \sin\theta_W \\
          -\sin\theta_W & \cos\theta_W
         \end{array}
  \right)
  \left(\begin{array}{c} B_0 \\ W_0 \end{array} \right)
  \label{eq:wzafields}
\end{equation}
for the charged $W$-bosons, $W^\pm$, the neutral $Z$-boson, $Z^0$, and 
the photon, $A^0$, respectively.  The weak mixing angle, $\theta_W$, relates 
the $SU(2)$ and $U(1)$ coupling constants to the electromagnetic coupling 
constant (\ie\ the fine structure constant), $\alpha$, by
\begin{equation}
  g^2=4\pi\alpha/\sin^2\theta_W, \: \:\: g^{\prime2}=4\pi\alpha/\cos^2\theta_W.
  \label{eq:thetaw}
\end{equation}
The gauge boson masses are given by
\begin{equation}
  \Mw = gv/2,\:\:\: \Mz = v\sqrt{g^2 + g^{\prime2}}/2, \:\:\: M_{A} = 0
  \label{eq:mwmzma}
\end{equation}
corresponding to the massive $W\pm$ and $Z^0$-bosons and the
massless photon, \linebreak respectively.  Equations~\ref{eq:thetaw} 
and \ref{eq:mwmzma} yield the following relationship, \linebreak 
$\sin^2\theta_W = 1 - (\Mw/\Mz)^2$.

At low energies, the electroweak theory is equivalent to the Fermi theory
of weak interactions.  Comparing the electroweak Lagrangian in 
Equation~\ref{eq:ewklagrangian} to Fermi's expression for the weak interaction
yields the following equality, \linebreak
$G_F = g^2/(4\sqrt{2}\Mw^2) = \pi\alpha/(\sqrt{2}\Mw^2\sin^2\theta_W)$.  
This can be rewritten as
\begin{equation}
  \Mw^2\left(1-\frac{\Mw^2}{\Mz^2}\right) = \frac{\pi\alpha}{\sqrt{2}G_{F}}
  \label{eqn:ewksmtlevel}
\end{equation}
relating the mass of the $W$-boson, the mass of the $Z$-boson, the fine 
structure constant and the Fermi constant, so that a measure of three yields 
a prediction of the fourth.  To obtain theoretical predictions of an precision
comparable to that of the experimental determinations of these parameters, 
radiative corrections must be included.  These corrections can  be incorporated
by rewriting Equation~\ref{eqn:ewksmtlevel} as:
\begin{equation}
  \Mw^2\left(1-\frac{\Mw^2}{\Mz^2}\right) = \frac{\pi\alpha}{\sqrt{2}G_{F}}
    \left(\frac{1}{1-\Delta r}\right)
  \label{eqn:ewksmrc}
\end{equation}
where the effects of the radiative corrections are included in the additional 
term, $\Delta r$.  The corrections can be separated into three
main pieces, 
\begin{equation}
  \Delta r = \Delta\alpha + \Delta\rho(\Mt^2) + \Delta\chi(\ln(\Mh/\Mz)),
\end{equation}
which include the running of the fine structure
constant, $\Delta\alpha$, a quadratic dependence on the top quark mass, 
$\Delta\rho$, and a logarithmic dependence on the mass of the Higgs boson,
$\Delta\chi$~\cite{rcreview}.   This last dependence is a unique 
consequence of the non-Abelian gauge structure of the electroweak theory, which
allows interactions among the gauge bosons themselves.  It is because of these
radiative corrections that precision measurements of $G_F$, $\alpha$, \Mz\ and
\Mw\, when compared with theoretical calculations, can yield 
constraints on \Mt\ and \Mh~\cite{sirlin}.

%
\section{MEASUREMENTS OF \Mw\ AT $\pp$ COLLIDERS}

\subsection {Measurement Techniques}

\subsubsection {$W$-BOSON PRODUCTION}

\begin{figure}[htb]
  \begin{center}
    \psfig{figure=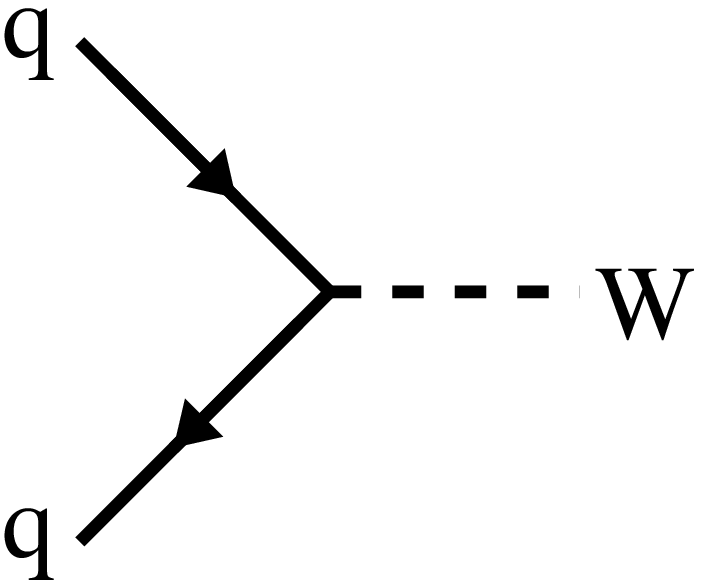,width=1.2in}
    \psfig{figure=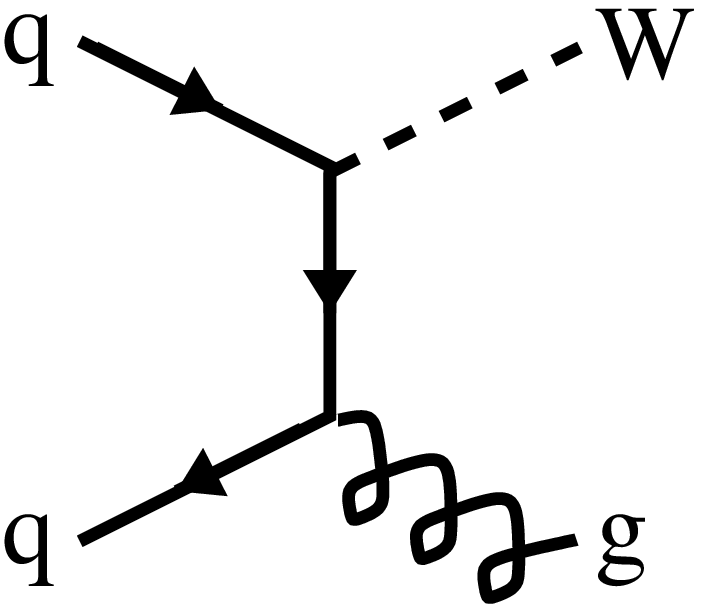,width=1.2in}
    \psfig{figure=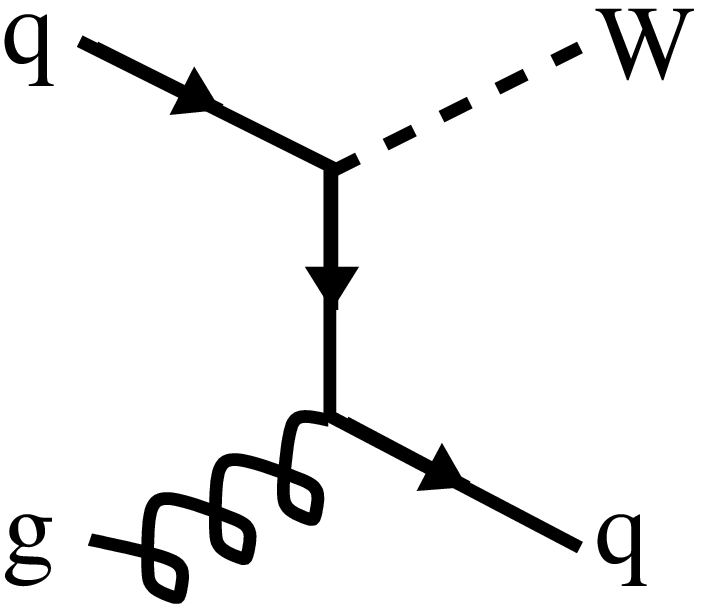,width=1.2in}
    \caption{Feynman Diagrams for $W$-boson production in $\pp$ collisions.}
    \label{fig:Wprod}
  \end{center}
\end{figure}

There have been two $\pp$ colliders with sufficient center-of-mass energy 
($\sqrt{s}$) to produce $W$-bosons, the \SppS\  at CERN ($\sqrt{s}$=630 GeV) 
and the Tevatron at Fermilab ($\sqrt{s}$=1.8 TeV). Figure \ref{fig:Wprod} shows
the most important subprocesses for $W$-boson production in $\pp$ collisions. 
At these center-of-mass energies, the dominant subprocess is $\qq\to W$. 
$Z$-bosons, which form an essential control sample, are produced via analogous
processes.

The $W$-boson mass measurements from these colliders all make use of the $\Wev$ and $\Wuv$ decay channels. Electrons and muons\footnote{Here and throughout
this article charge conjugation is implied.} are easy to trigger on and their momenta can be measured very precisely. Moreover, $W$ and $Z$-bosons are the dominant source of isolated, high-$p_T$ electrons and muons in $\pp$ collisions. Therefore, samples of $W$ and $Z$-decays involving electrons and muons can be identified with very little background. Purely hadronic decays of the $W$-boson are swamped by QCD background. Decays involving $\tau$ leptons are difficult to identify because the $\tau$ leptons decay before they enter the detector. 

The cross sections for $W$ and $Z$ production in $\pp$ collisions 
are large, $\sigma\times{\mathcal{B}}=680$ pb at $\sqrt{s}$=630 GeV 
\cite{UA2-Wprod} and 2.3 nb at $\sqrt{s}$=1.8 TeV \cite{D0-Wprod} for 
$W$-bosons, where $\mathcal{B}$ is the leptonic branching fraction. For 
$Z$-bosons the corresponding values are about 10 times smaller. 

In the following, we refer to a coordinate system which has its origin at the 
average $\pp$-collision point. The $z$-axis is defined by the proton 
beam. The $y$-axis points up. The $x$-axis points along the horizontal. 
Since the parton center-of-mass frame is boosted along the beam direction, 
momentum components transverse to the beam are especially important. They are
are denoted by a subscript $_T$. The beams are unpolarized so that there is an 
inherent azimuthal symmetry. Thus it is often convenient to work in a cylindrical coordinate system in which $\phi$ is the angle with the $x$-axis in the $x$-$y$ plane.
The longitudinal phase space is most conveniently expressed in terms of the 
pseudo-rapidity, $\eta=-\ln\tan\left(\theta/2\right)$, which is related to the polar angle $\theta$.

The detectors have approximate azimuthal and forward-backward symmetry. They are constructed to cover as large a region in pseudo-rapidity as possible. From inside out, they typically consist of several sub-detectors: a tracking system to measure the trajectories of charged particles; a calorimeter to measure the energy of electrons, photons, and hadrons; and a muon detection system. The tracking system may be located in a magnetic field to determine the momentum of charged particles from the curvature of their trajectories.

\subsubsection {EVENT CHARACTERISTICS}
 
The detectors register the charged lepton from the decay of the $W$-boson, while the neutrino escapes without detection. 
The initial proton and antiproton break up in the collision and the fragments 
hadronize, contributing additional particles to the event. The hadronization 
of final-state quarks or gluons also contributes particles, which may form 
jets if the initial parton had sufficient transverse momentum. We refer to all 
particles, except the $W$-boson and its decay products, as the underlying event. 
Some of the particles of the underlying event escape through the beam pipe and 
are not detected at all. These particles may carry a substantial momentum component along the beam axis, but they carry little momentum transverse to the beam.

The transverse momenta of all final-state particles must add to zero, because the initial $\pp$ momentum is zero and momentum is conserved. Since the undetected neutrino carries away substantial momentum, the transverse momenta of all observed final-state particles do not add to zero. The apparent transverse  
momentum imbalance is called ``missing $\pT$''. 

The particles of the underlying event that fall within the detector acceptance 
cannot all be detected individually. The detector measures the sum of the energy 
of all particles incident on one calorimeter segment. The quantity
\begin{equation}
\vuT = \sum_i E_i\sin\theta_i \hat\imath
\end{equation}
gives an approximate measurement of the total transverse momentum of all 
underlying event particles. The sum runs over all calorimeter cells, except those assigned to the charged lepton. $E_i$ is the energy in cell $i$. The unit vector $\hat\imath$ forms a right angle with the beam axis and points from the beam to cell $i$. 

Thus, the basic observables are the momentum of the charged lepton ($\vp^\ell$, $\ell$=$e$ or $\mu$) and the sum of the transverse momenta of the particles in the underlying event ($\vuT$) which we call the recoil momentum. From these, the transverse momenta of the $W$-boson ($\vpT^W = - \vuT $) and the neutrino ($\vpT^\nu= - \vuT - \vpT^\ell$) can be inferred.
A high-$\pT$ charged lepton and large missing $\pT$ form the characteristic 
signature of $W$-boson decay events. $Z$-decay events are characterized by two charged leptons with high $\pT$. There are no high-$\pT$ neutrinos in such 
$Z$-decays and therefore no significant missing $\pT$ is expected. 

\subsubsection {MASS MEASUREMENT METHOD}

It is not possible to reconstruct the invariant mass of the $W$-boson because there is no measurement of the momentum component of the neutrino along the beam axis. In addition, the $W$-bosons are neither produced at rest nor are they the only particles produced in the collisions. Therefore, a precision measurement of the $W$-boson mass using $W$-bosons produced in $\pp$ collisions poses a particular challenge.

The most precise measurements of the mass of the $W$-boson are based on the transverse mass of the charged lepton-neutrino pair, defined as
\begin{equation}
\mT = \sqrt{2\pT^\ell\pT^\nu\left(1-\cos\left(\phi^\ell-\phi^\nu\right)\right)}.
\end{equation}
The advantage of the transverse mass is its invariance under boosts along the beam axis. Boosts transverse to the beam axis only give rise to corrections of order $(\pT^W/E^W)^2$. On the other hand, the transverse mass depends on the measurement of the recoil momentum $\vuT$ and all the associated systematic effects.

An alternative method to determine \Mw\ uses the $\pT$ spectrum of the lepton. This has the advantage of being insensitive to $\uT$. However, it is affected by the boost of the $W$-boson transverse to the beam axis to order $\pT^W/E^W$ and is therefore much more sensitive to systematic effects associated with the production of the $W$-bosons.  

In principle, the charged lepton momentum or the transverse momentum of the neutrino can also be used to measure the $W$-boson mass. However the former is sensitive to boosts in all directions and the latter suffers from poor resolution. These variables serve mainly as cross checks.

It is not possible to describe the spectra of the variables mentioned above analytically. They have to be calculated numerically using a Monte Carlo model that takes into account the mechanisms for production and decay of $W$-bosons, and detector effects.

\subsubsection {BACKGROUNDS}

The backgrounds to the $\Wlv$ signal are 
$W\to\tau\overline{\nu}\to\ell\overline{\nu}\overline{\nu}\nu$ (1-2\%), hadronic backgrounds (1\% for $\Wev$, $\ll$1\% for $\Wuv$), $\Zll$ ($\ll$1\% for $\Wev$, 4\% for $\Wuv$), and cosmic rays ($\ll$1\%). Hadronic backgrounds arise hadrons, that fake the charged lepton signature. $\Zll$ decays can enter the $W$ sample if one of the leptons escapes detection. The quoted percentages give the approximate residual fractions of background events in the final $W$ samples. The precise background contamination depends on the details of the event selection and the detector. They have to be taken into account in the measurement to avoid biasing the result.
The normalization and shapes of the background spectra are determined from control data samples and detailed Monte Carlo simulations.

\subsubsection {EVENT SELECTION}

The event selection consists of the identification of the charged lepton and a set of kinematic and topological cuts. The selection criteria have to achieve two competing goals: to reject backgrounds efficiently and to minimize any biases to the selected event sample. Kinematic cuts, requiring the momentum of the charged lepton and missing $\pT$ above a threshold (typically 25 GeV), are easy to simulate and reduce backgrounds significantly. $W$-bosons with very high transverse momenta do not add to the statistical significance of the mass measurement, because their transverse mass and lepton $\pT$ spectra are very broad and carry little mass information. In addition, their recoil response is difficult to simulate and they are subject to higher background contamination. Thus, such events are usually eliminated from the sample by requiring that the $W$-boson $\pT$ is below some threshold and/or that there are no high-$\pT$ jets in the events. 

An electron is typically identified as an energy deposit in the calorimeter, consistent in shape with an electromagnetic shower, and a track that points to it. Since these electrons are highly relativistic, their momenta can be equated to the energy measured in the calorimeter.
A muon is typically identified as a track stub in the muon detection system that matches a track in the tracking system and energy deposits in the calorimeter, small enough to be consistent with the passage of a minimum-ionizing particle. These criteria reduce contamination from hadronic backgrounds. However, both criteria inherently require the lepton to be isolated from other activity in the event. This biases the selection towards event topologies in which the charged lepton is emitted along the direction of motion of the $W$-boson. In such events, the probability is smaller that the lepton overlaps with a recoil particle. Since the boost of the $W$-boson increases the lepton $\pT$ in the lab frame, these events tend to have harder charged leptons and softer neutrinos. This bias does not affect the transverse mass 
spectrum significantly, but it must be understood to predict the $\pT$ spectra of the charged leptons and the neutrinos correctly.

Specific cuts are required to reject events due to an accidental coincidence between a cosmic ray traversing the detector and a beam crossing. 

\subsubsection{MONTE CARLO MODEL}

In this section a generic description of the Monte Carlo models is given. The 
sections below, describing the individual measurements, highlight significant 
experiment specific deviations. 
To keep statistical fluctuations in the Monte Carlo simulation negligible, many millions of $W$-decay events have to be generated. To simulate such large event samples, parameterized Monte Carlo algorithms for $W$-boson production and decay, and detector modeling were developed specifically for the $W$-mass measurements. 

First, the $W$-bosons are generated.
Their $\pT$ distribution is determined theoretically from QCD-based calculations, empirically from the observed $\pT$ distribution of $Z$-bosons, or by a combination of both. 
The rapidity distribution of the generated $W$-bosons depends on the momentum distribution of the partons inside the proton. To determine it, a specific set of parton distribution functions must be chosen.
The mass distribution of the generated $W$-bosons is a relativistic Breit-Wigner curve with peak at the hypothesized value of the $W$-boson mass and $s$-dependent width, given by the Standard Model expectation. This mass spectrum is skewed towards lower mass values due to the momentum distribution of the partons inside the proton.

Next, the $W$-bosons are allowed to decay. At lowest order, the angular distribution of the charged leptons is ${d\sigma/d\cos\theta^*} \propto \left(1-\xi q \cos\theta^*\right)^2$,
where $\theta^*$ is the scattering angle of the charged lepton in the rest frame of the $W$-boson, $q$ the charge of the lepton, and $\xi$ the helicity of the $W$-boson. In most events the initial quark comes from the proton and $\xi$ equals $-1$. In the much less likely case that the initial antiquark comes from the proton $\xi$ equals $+1$. Higher-order QCD processes modify the angular distribution of the leptons. Radiative decays, in which $\Wlv\gamma$, modify the momentum spectrum of the leptons. The Monte Carlo models either include these effects or corrections are applied to the results.

The decay $W\to\tau\overline\nu\to \ell\overline\nu\overline\nu\nu$ leads to events, that are topologically indistinguishable from $\Wlv$. These can be calculated precisely in the framework of the Standard Model and are typically included in the Monte Carlo model.

Finally, the Monte Carlo model must account for detector effects.
The simulation has now generated the ``true" momenta of the $W$-boson (and thus the recoil momentum) and the charged lepton in the event. These are modified to account for experimental resolutions, biases, and efficiencies. Adding random Gaussian uncertainties to the observables simulates resolution effects. The widths of these Gaussian distributions are parameterized in detector-specific ways. Other effects accounted for include the response of the detector to the charged lepton and to the underlying event. Also modeled are the imperfect separation of energy deposits between lepton and underlying event, which leads to biases in lepton and recoil momentum measurements, and selection efficiencies that depend on kinematics or topology of the events. 

Events due to the process $\pp\to Z+X$, $\Zll$ constitute an extremely important control sample to determine these effects. Comparing the observed $Z$ peak to the known $Z$-boson mass calibrates the energy or momentum response to charged leptons. The observed width of the $Z$ peak measures the energy or momentum resolution for charged leptons. The $Z$-boson $\pT$ can be measured directly using the charged leptons from its decay and indirectly from the recoil momentum. By comparing both determinations, the $Z$ events also serve to calibrate the recoil momentum response of the detector relative to the charged lepton response. 

\subsubsection {MASS MEASUREMENT}

The Monte Carlo model predicts the shape of the transverse mass and the lepton $\pT$ spectra from $\Wlv$ decays as a function of the hypothesized value of the $W$-boson mass. These are added to the estimated background spectra and normalized to obtain probability density functions for a maximum likelihood fit to the spectra from the collider data. The statistical uncertainty in the fit to the $\mT$ spectrum is typically 11 GeV/(number of events)$^{1/2}$.

\begin{figure}[htp!]
  \begin{center}
    \psfig{figure=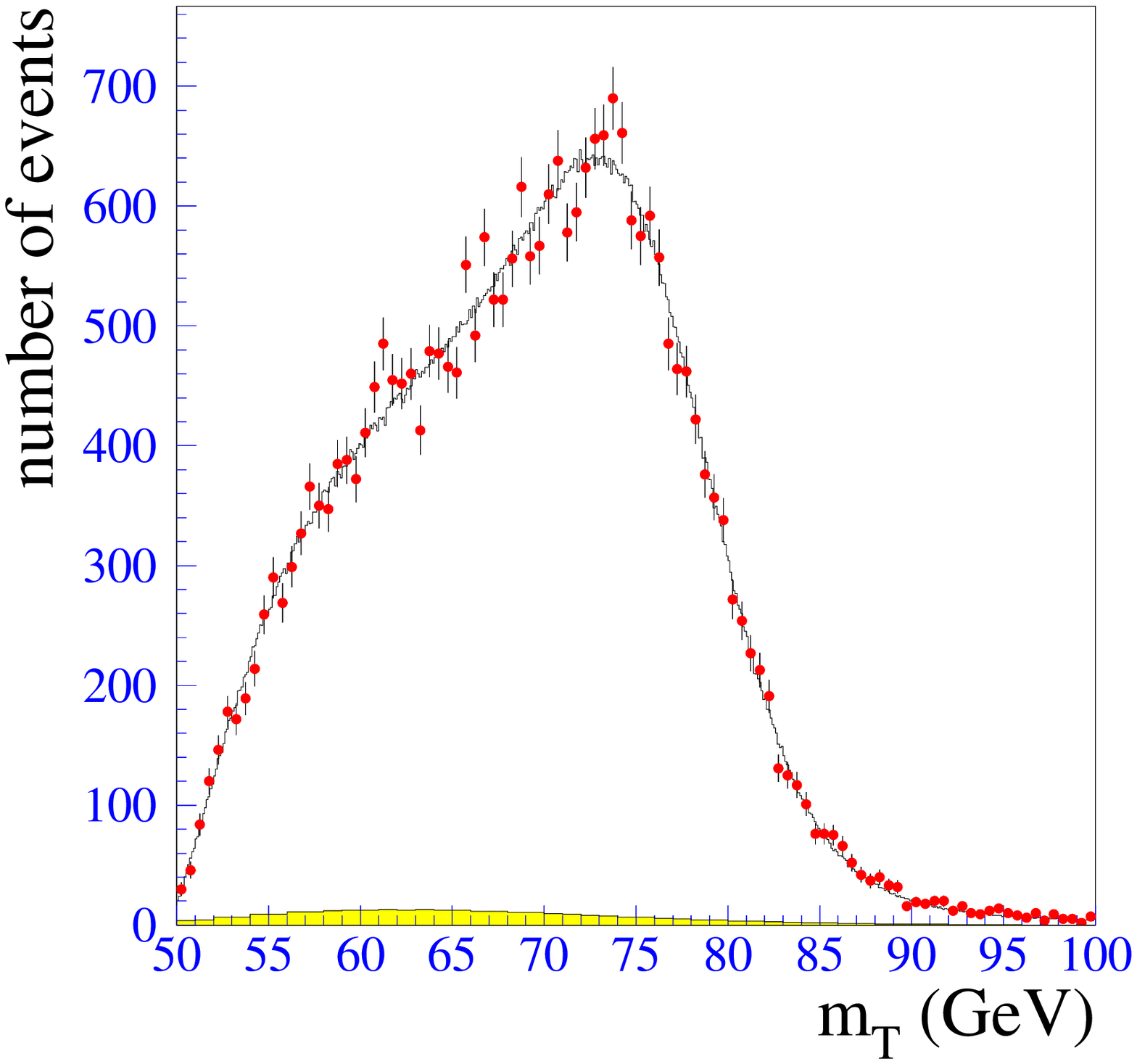,width=4in}
    \vspace{-0.25in}
    \caption{Transverse-mass spectrum from $W$-decays measured by 
      \Dz\ \cite{D0_Ib}.}
    \label{fig:mT}
    \psfig{figure=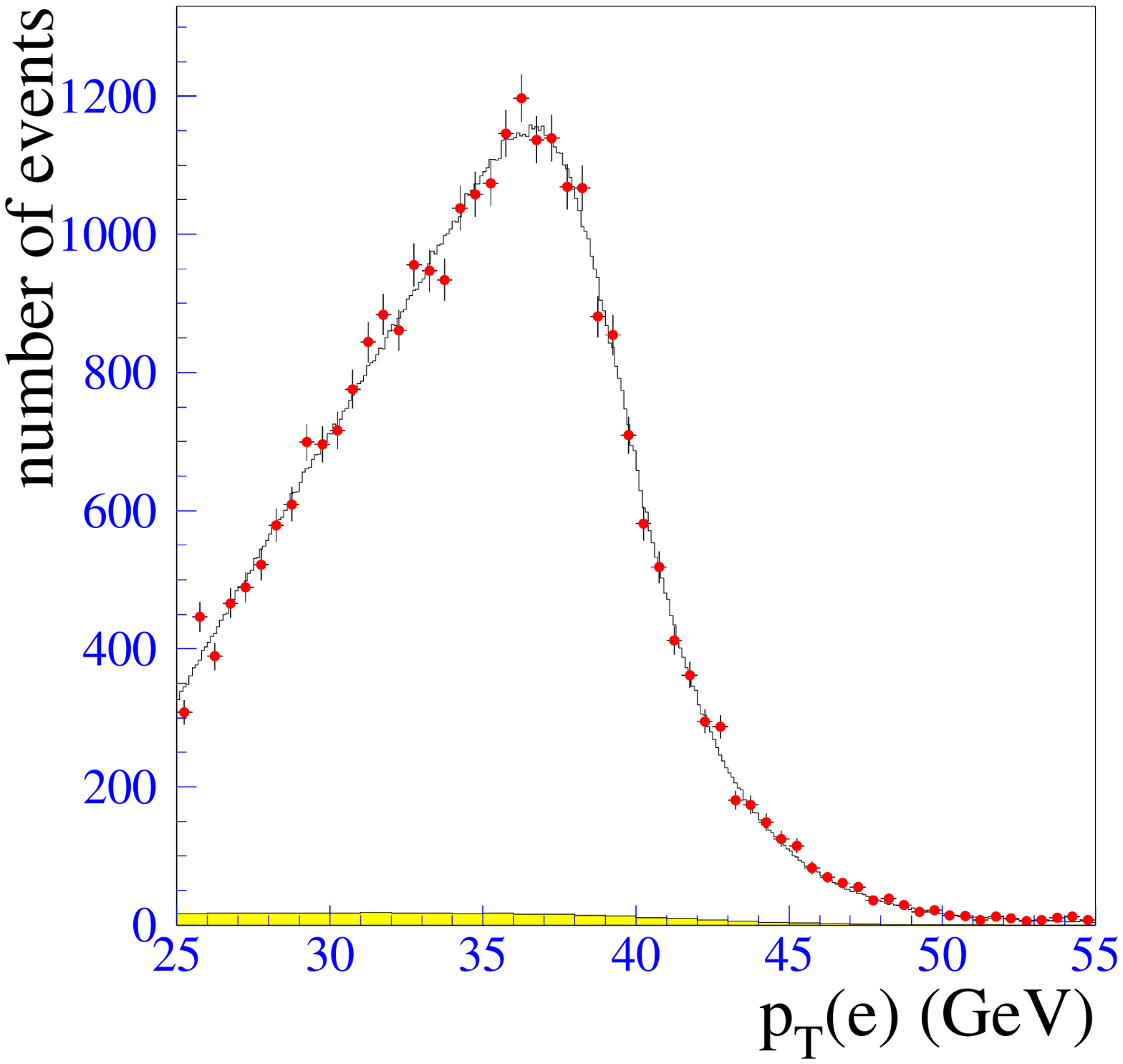,width=4in}
    \vspace{-0.25in}
    \caption{Electron $\pT$ spectrum from $W$-decays measured by 
      \Dz\ \cite{D0_Ib}.}
    \label{fig:pT}
  \end{center}
\end{figure}

Figures \ref{fig:mT} and \ref{fig:pT} show representative spectra of transverse mass and lepton $\pT$. These spectra were measured by the \Dz\ experiment. 
The points indicate the collider data, the line indicates the Monte Carlo prediction that agrees best with the data, and the shaded region indicates the estimated background contribution. 

\subsubsection{SYSTEMATIC UNCERTAINTIES}

All inputs of the Monte Carlo model contribute to the systematic uncertainty of
 the measurement. Each such contribution is estimated by varying the input 
within its 68\% confidence interval to determine the resulting change in the 
measured $W$-boson mass. The total systematic uncertainty is the sum in 
quadrature of all such contributions. Most of the model parameters are 
constrained by control data samples, most notably by the $\Zll$ samples. In 
most cases the precision with which these parameters can be determined is 
limited by the size of the control samples, so that these uncertainties are 
really statistical in nature. This means that they can be quantified in a 
well-defined way. There are some cases in which no rigorous confidence interval
can be defined, as is usually more typical of systematic uncertainties.

The following paragraphs elaborate on the most important categories of 
systematic uncertainties. The values quoted for each uncertainty are typical of
the measurements from the Tevatron using a fit to the $\mT$ spectrum from a 
data sample of about 100 pb$^{-1}$. 

\begin{description}

\item[Lepton energy/momentum scale:] (70-85 MeV) This is the most important 
systematic effect. Since all detector responses are calibrated against the 
charged leptons using the $Z$ sample, the measured $W$ mass simply scales with
the lepton scale. It can be set in two ways. 

One method is to calibrate the lepton scale so that the $\Zee$ and $\Zuu$ mass
 peaks (Fig. \ref{fig:Zee}) agree with the known $Z$-boson mass \cite{Z_mass}.
 This has the advantage that the uncertainty is dominated by statistical 
fluctuations in the $Z$ sample, $\approx$3 GeV/(number of events)$^{1/2}$, and
 little extrapolation is needed to the energies of leptons from $W$-decays. 
Uncertainties in the extrapolation can be limited by using other resonances, 
such as $J/\psi\to\epem$ or $\pi^0\to\gamma\gamma$. If the scale calibration of
 the charged leptons is tied to the $Z$-boson mass, the measured quantity is 
really the ratio of the $W$ and $Z$-boson masses, rather than the $W$-boson 
mass. Given that the $Z$-boson mass is known so precisely, the two 
quantities are of course de facto equivalent.

\begin{figure}[htp!]
  \begin{center}
    \psfig{figure=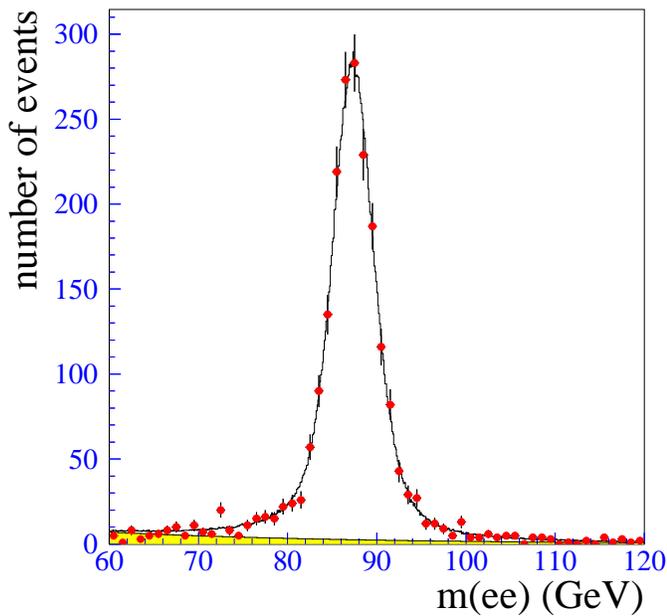,width=4in}
    \vspace{-0.25in}
    \caption{Mass spectrum from $\Zee$ decays measured by \Dz\ \cite{D0_Ib}.}
    \label{fig:Zee}
  \end{center}
\end{figure}

Detectors with a magnetic tracking system can alternatively calibrate the 
momentum measurement for charged tracks, using for example 
$J/\psi\to\mu^+\mu^-$ decays (Fig. \ref{fig:Jpsi}), and then extrapolate to the
 momentum of leptons from $W$-decays. The calorimeter must then be calibrated 
against the track momentum using the ratios of energy and momentum ($E/p$) 
measured for electrons from $W$-decays. The advantage of the latter method lies
 in the higher precision of the track momentum calibration. Its disadvantage 
are the systematic effects associated with the extrapolation to higher momenta
 and the effects of radiation on the $E/p$ spectrum.

\begin{figure}[htp!]
  \begin{center}
    \psfig{figure=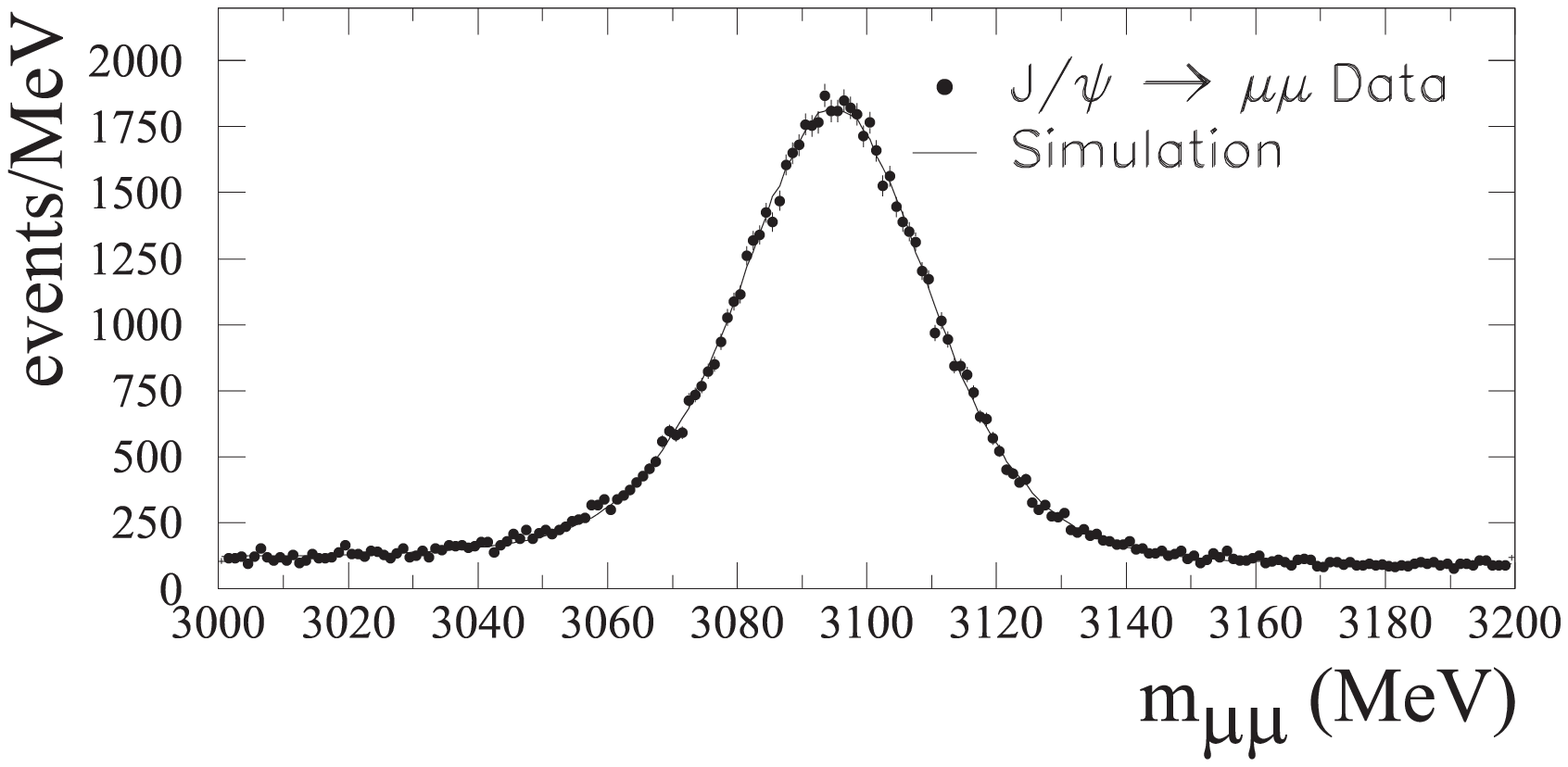,width=4in}
    \vspace{-0.25in}
    \caption{Mass spectrum from $J/\psi\to\mpmm$ decays measured by 
      \CDF\ \cite{CDF-95}.}
    \label{fig:Jpsi}
  \end{center}
\end{figure}

\item[Lepton energy/momentum resolution:] (20-25 MeV) The electron energy 
resolution can be modeled as 
$\sigma/E=({\mathcal{S}}^{2}/E+{\mathcal{C}}^{2})^{1/2}$, where $E$ is the 
electron energy, $\mathcal{S}$ the sampling term and $\mathcal{C}$ the constant
 term. The value of $\mathcal{S}$ is taken from beam tests and $\mathcal{C}$ is
 chosen so that the width of the $Z$ peak predicted by the Monte Carlo model 
agrees with the $Z$ peak from collider data. For muons, the transverse momentum
 resolution is of the form $\sigma/\pT^2=\kappa$, where $\pT$ is the transverse
 momentum of the muon and $\kappa$ a constant chosen to match the widths of the
 $Z$ peaks from the Monte Carlo model and the collider data.

\item[Recoil model:] (30-40 MeV) This uncertainty arises from the parameters describing the response and resolution of the detector to the underlying event. These are determined from $Z$-decays and to a lesser extent from $W$-decays. 

\item[Lepton removal:] ($\approx$ 15 MeV) This describes uncertainties in the 
corrections to the recoil momentum for the imperfect separation of energy 
deposits between the charged lepton and underlying event. Some particles from 
the underlying event inevitably overlap with the charged lepton in the 
calorimeter. Their energies are not included in the calculation of $\uT$. The 
correction is equal to the average energy deposited by the underlying event in
 an appropriately sized calorimeter segment in the $W$ data sample. 

\item[Proton structure:] (10-20 MeV) This uncertainty characterizes the variations in the result between different choices of parton distribution functions. While relatively small, this uncertainty is completely correlated for all measurements at $\pp$ colliders. For any given set of parton distribution functions, the variation in the measured $W$-boson mass is strongly correlated with the variation in the predicted forward-backward charge asymmetry in $\pp\to\Wlv$ \cite{CDF-95}. Recently, increasingly precise measurements of this asymmetry by the \CDF\ collaboration \cite{CDF-Wasym_Ia, CDF-Wasym_I} have helped constrain the parton distribution functions (e.g. \cite{MRSA, CTEQ}) and reduce the resulting uncertainty in the $W$-boson mass measurement. The agreement of recent parton distribution functions with the measured asymmetry is shown in Fig. \ref{fig:Wasym}.
Since no complete error matrices are available for parton distribution functions, this uncertainty cannot be evaluated in a statistically rigorous fashion. 

\begin{figure}[htb!]
  \begin{center}
    \psfig{figure=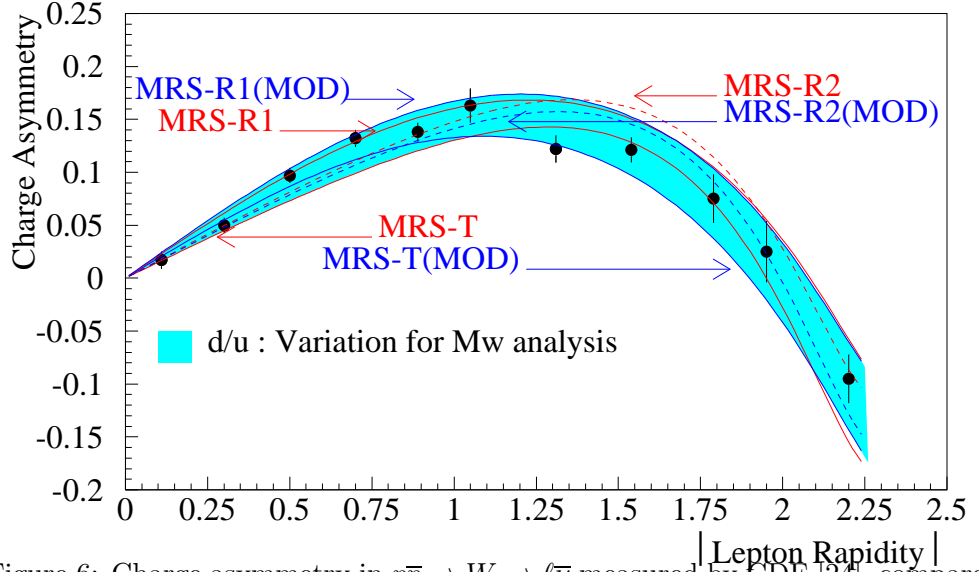,width=5in,height=3in}
    \vspace{-0.25in}
    \caption{Charge asymmetry in $\pp\to\Wlv$ measured by \CDF\ 
      \cite{CDF-Wasym_I}, compared to predictions of different parton 
      distribution functions.}
    \label{fig:Wasym}
  \end{center}
\end{figure}

\item[$W$ $\pT$ spectrum:] ($\approx$ 10 MeV) The shape of the $W$ $\pT$ distribution can be constrained by theoretical calculations in conjunction with the observed $Z$ $\pT$ distribution. For low $\pT$, the $W$ $\pT$ spectrum cannot be calculated perturbatively. One approach \cite{LY, AK} uses the Collins-Soper-Sterman resummation formalism \cite{CSS,AEMG} which contains universal empirical parameters. These parameters can be constrained by requiring the calculation to agree with the observed $Z$ $\pT$ spectrum. Another, essentially equivalent, approach is to use the observed $Z$ $\pT$ spectrum directly and convert it to a prediction for the $W$ $\pT$ spectrum using the ratio of the calculated $W$ and $Z$ $\pT$ spectra. The precision of both approaches is limited by statistical fluctuations in the $Z$ data sample. The measured $W$ $\pT$ spectrum does not provide a very stringent constraint because its shape at low $\pT$ is dominated by the recoil response of the detector. The $Z$ $\pT$ spectrum, on the other hand, can be measured independently of the recoil response using the two charged leptons from the decay of the $Z$. 

\item[Higher-order corrections:] (10-20 MeV) 
Calculations of radiative $W$-boson decays involving one photon \cite{W_rad-BK, W_rad-Baur} and two photons \cite{W_rad2-Baur} are available. The precision of these corrections is limited by experimental factors which determine whether the photons and the lepton are resolved. 

\item[Backgrounds:] (5-25 MeV) The hadronic background dominates this uncertainty for the $\Wev$ channel. Normalization and shape are determined from control data samples. The $\Zuu$ background dominates the $\Wuv$ channel. Uncertainties in the tracking efficiency at high $\left|\eta\right|$ and in the parton distribution functions give rise to this uncertainty.

\end{description}

\subsection {Individual Measurements}

\subsubsection {\UAT\ EXPERIMENT}

The \UAT\ collaboration published the first measurement of the $W$-boson mass with a precision below 1\% \cite{UA2-90}. This was superceded by an improved result \cite{UA2-92} based on 13 pb$^{-1}$ of data taken in 1988-1990 at the CERN \SppS\ collider at $\sqrt{s}$=630 GeV. 

The \UAT\ detector \cite{UA2} consists of a calorimeter which covers the pseudo-rapidity range $|\eta|<3$. It consists of lead and iron absorber plates interspersed with scintillators and wavelength shifter readout. The electromagnetic section is 17-24 radiation lengths deep and segmented into elements covering 15$^\circ$ in azimuth and approximately 0.2 units in pseudo-rapidity. The electron energy resolution is $\sigma/E=17\%/\sqrt{E/\mbox{GeV}}$. 
The hadronic section is four interaction lengths deep.
Inside the calorimeter are nested cylindrical tracking detectors. From inside out, they are: a drift chamber with arrays of silicon pad counters on either side, a transition radiation detector, and a scintillating fiber detector. The detector has no magnetic field.

The $W$-boson mass measurement uses the $\Wev$ and $\Zee$ decay channels. 
The selection for the $W$-event sample requires an electron in the central calorimeter, $\pT^e>$20 GeV, $\pT^\nu>$20 GeV, $\uT<20$ GeV, and $40<\mT<120$ GeV. The Monte Carlo model calculates the $W$-boson rapidity from HRMSB structure functions \cite{HMRSB}. The spectrum of $\pT^W$ is taken from a calculation \cite{AK}, modified by an empirical distortion function. The distortion function is determined by comparing the spectrum of $\pT^Z$ predicted by the same calculation with the observed  $\pT^Z$ distribution. The recoil response model has two parameters: resolution (dependent on the total energy measured in the event), and offset (dependent on $\pT^W$). Both were tuned using the $\Zee$ sample and requiring that the mean $\pT^W$ predicted by the model agrees with the data. 

Two $Z$ samples are used. Sample 1 requires two central electrons, which must be inside the fiducial volume of the calorimeter within $|\eta|<0.8$. Sample 2 requires one central electron and one electron outside the central acceptance region. The energy of the ``outside'' electron is rescaled so that all transverse momentum components along the outer bisector of the two electron directions add to zero. 

A fit to the transverse mass spectrum gives $80.84\pm0.22\pm0.83$ GeV\footnote{Whenever two uncertainties are given, the first is due to statistical fluctuation, the second to systematic effects.}. The $Z$ mass is measured to be $91.74\pm0.28\pm0.93$ GeV using both $Z$ samples. In all cases, the systematic uncertainties are dominated by the electron energy scale calibration.
In the ratio $\Mw/\Mz$ the energy scale and other systematic uncertainties partially cancel. \UAT\ finds $\Mw/\Mz$=0.8813$\pm$0.0036$\pm$0.0019.
Using the current $Z$ mass of 91.187$\pm$0.002 GeV \cite{Z_mass} gives \footnote{Updated relative to original publication.} $\Mw=80.36\pm0.33\pm0.17$ GeV.

\subsubsection {\CDF\ EXPERIMENT}

The \CDF\ collaboration has performed measurements of the $W$-boson mass using data sets from three running periods of the Fermilab Tevatron: 1988/89 \cite{CDF-90}, 1992/93 \cite{CDF-95}, and 1994-96 \cite{CDF-99}. A publication of the results from the 1994-96 data is in preparation. 

The \CDF\ detector \cite{CDF, CDF-Upgrades} is a multipurpose magnetic spectrometer. Tracking detectors are surrounded by a solenoid, that provides an axial magnetic field of 1.4 T. The vertex time-projection chamber measures the position of the $\pp$-collision point along the $z$-axis with 1 mm resolution. The central tracking chamber has 84 layers of wires and covers $40^\circ<\theta<140^\circ$. The transverse momentum resolution is $\sigma/\pT^2 = 0.0011/\mbox{GeV}$. 

The central calorimeter covers $|\eta|<1.1$. The electromagnetic section consists of lead plates interleaved with scintillator. Including chamber wall and solenoid, it is 19 radiation lengths deep and segmented into projective towers covering $\Delta\phi\times\Delta\eta$=15$^\circ\times$0.1. The electron energy resolution is $\sigma=13.5\%\sqrt{E\sin\theta/\mbox{GeV}}$. Proportional chambers after 6 radiation lengths measure the shower centroid position to 3 mm. The hadron calorimeter is made of iron-scintillator shower counters. Outside the central region ($1.1<|\eta|<4.2$) the calorimeter is made of gas proportional chambers with cathode pad readout.

Muon chambers are located 3.5 m from the beam behind 5 nuclear absorption lengths and cover $|\eta|<0.6$. 

\CDF\ use both the $\Wev$ and $\Wuv$ channels. Events are selected with $\pT^\ell>25$ GeV and $\pT^\nu>25$ GeV. There must be no high-$\pT$ tracks or energetic clusters in the calorimeter in addition to the charged lepton. For the 1992/93 data sample, $\uT<20$ GeV is required.

The muon momentum scale is based on a calibration of the tracking system to the $J/\psi$ mass. The electron energy scale is set using $E/p$ for electrons from $W$-decays. The calibration is checked using the $Z$ mass from $\Zee$ decays, 91.12$\pm$0.52 GeV (1988/89) and $90.88\pm0.19\pm0.20$ GeV (1992/93), using the same calibration as for $\Wev$ events.

For the analysis of the 1994-96 data, the electron energy scale determined by the $E/p$ technique results in a $Z$-mass peak from $\Zee$ decays 3.9 standard deviations below the known $Z$ mass. Thus this technique is not used to determine the $W$ mass. Instead, the muon momentum and electron energy scales are calibrated using the observed $Z$-mass peaks.

For the analysis of the 1988/89 data, the Monte Carlo model uses MRS-B parton distribution functions \cite{MRSB} as the nominal choice. The transverse momentum distribution of the $W$-bosons is obtained from the observed $\pT^W$ distribution by an unfolding procedure. The results from the fits to the $\mT$ spectra in both decay channels are listed in Table \ref{tab:summary}. Both combined give $\Mw = 79.91\pm0.39$ GeV.

For the analysis of the 1992/93 data, parton distribution functions are restricted to those consistent with the measured charge asymmetry in $\Wlv$ decays \cite{CDF-Wasym_Ia}. The Monte Carlo model uses MRSD$-$' \cite{MRSD} as the nominal choice. The transverse momentum distribution of the $W$-bosons is obtained from the observed $\pT^Z$ distribution, corrected for electron resolution and scaled so that the spectrum of the component of $\vuT$ perpendicular to the direction of the charged lepton agrees with the $W$ data. The underlying event model uses a lookup table of $\vuT$ versus generated $\vpT^W$, built from the $\Zee$ event sample. The results from the fits to the $\mT$ spectra in both decay channels are listed in Table \ref{tab:summary}. They combine to $\Mw = 80.41\pm0.18$ GeV.

For the analysis of the 1994/95 data, MRS-R2 \cite{MRSR2} parton distribution functions are used. The $\pT^W$ spectrum is derived from the observed $\pT^Z$ spectrum, corrected based on a theoretical calculation of the ratio of the $\pT^W$ and $\pT^Z$ spectra \cite{LY, AK}. The parameterized recoil model is tuned to $W$ and $Z$ data. The results from the fits to the $\mT$ spectra in both decay channels are listed in Table \ref{tab:summary}. Their combined value is $\Mw = 80.470\pm0.089$ GeV.

All \CDF\ measurements combined give $\Mw = 80.433\pm0.079$ GeV.

\subsubsection {\Dz\ EXPERIMENT}
The \Dz\ collaboration has published three measurements of the $W$-boson mass using the $\Wev$ channel. Two measurements, using data from 1992/93 \cite{D0_Ia} and 1994-96 \cite{D0_Ib}, use electrons in the central calorimeter. The third uses data with the electron in the end calorimeters \cite{D0_EC}.

The \Dz\ detector \cite{D0} is based on a hermetic uranium-liquid argon sampling calorimeter, which encloses a non-magnetic tracking system and is surrounded by a muon spectrometer. 

The tracking system consists of nested cylindrical sub-detectors: a vertex drift chamber, a transition radiation detector, and a central drift chamber, covering the pseudo-rapidity region $|\eta|<1$. Forward drift chambers on either side extend the tracking coverage to $|\eta|<3$. The chambers provide measurements of direction and energy loss of charged particles. 

The calorimeter is housed in three cryostats. The central calorimeter (CC) covers $|\eta|<1$ and the two end calorimeters (EC) cover $1<|\eta|<4$. The electromagnetic section is 21 radiation lengths deep and segmented radially into four layers and laterally into towers covering $\Delta\phi\times\Delta\eta$=0.1$\times$0.1. It measures the energy of electromagnetic showers with a resolution of $\sigma/E=13.5\%/\sqrt{E\sin\theta/\mbox{GeV}}$ and the shower centroid position with a resolution of 2.5 mm in azimuthal direction. The hadron calorimeter is 7-9 nuclear interaction lengths deep and provides hermetic coverage without projective cracks. 

The event selection for $W$-decay events requires $\pT^e>25$ GeV, $\pT^\nu>25$ GeV, and $\uT<15$ GeV. 

The $W$-boson $\pT$ and rapidity spectra are determined by a theoretical calculation \cite{LY}, constrained against the observed $\pT^Z$ spectrum, and the MRSA' parton distribution functions \cite{MRSA} for the 1992/93 data, and the MRST parton distribution functions \cite{MRST} for the 1994-96 data. 
The electron energy scale calibration is mainly based on the observed $Z$ peak. The energy spread of electrons from $Z$-decays, and signals from $J/\psi$ and $\pi^0$ decays limit nonlinearities. Electron resolution and the recoil model parameters are determined from the $Z$ data.

Based on the fit to the $\mT$ spectrum from the 1992/93 data, the \Dz\ collaboration measures\footnote{Updated uncertainties \cite{D0_Ib} relative to original publication.} $\Mw = 80.35\pm0.21\pm0.15$ GeV.
Based on the 1994-96 data, the \Dz\ collaboration measures the $W$-boson mass using the $\mT$, $\pT^e$, and $\pT^\nu$ spectra for electrons in CC and EC. 
Using the complete $6\times6$ covariance matrix, these results are combined to $\Mw=80.498\pm0.095$ GeV with $\chi^2=5.1$ for five degrees of freedom. By increasing the acceptance for electrons to pseudo-rapidity between $-$2.5 and 2.5, the sensitivity to the rapidity spectrum of the $W$-bosons is greatly reduced. This is reflected in the reduced uncertainty due to proton structure. This uncertainty is 15 MeV if only central electrons are included and 7 MeV if also electrons in EC are accepted.

All \Dz\ measurements combined give $\Mw = 80.482\pm0.091$ GeV.

\subsection {Combination of $\pp$ Collider Results}

Table \ref{tab:summary}\ lists the individual measurements for comparison in sequence of their publication. The number of $W$-boson events given reflect the number of events included in the fit to the transverse mass spectrum. The number of $Z$ events is given if the $Z$ data were used to calibrate the lepton scale. The statistical uncertainty reflects statistical fluctuations in the $W$ data sample. The scale uncertainty refers to the uncertainty in the lepton momentum scale calibration. If the $Z$ data are used to calibrate the lepton scale, this uncertainty is dominated by statistical fluctuations in the $Z$ data sample. The systematic uncertainty reflects all other systematic effects. All uncertainties are rounded to the nearest 5 MeV.

\begin{table}[ht!]
\begin{center}
\caption{Comparison of individual \Mw\ measurements from $\pp$ colliders}
\label{tab:summary}
\begin{tabular}{@{}l@{}c@{}crrrrrrr@{}}
\hline\hline
\multicolumn{3}{l}{measurement} & \multicolumn{2}{c}{events} & $\Mw$ & stat & scale  & syst & total \\
     &       &         & $W$ & $Z$ & (GeV) & \multicolumn{4}{c}{(MeV)} \\
\hline
\UAT & \cite{UA2-90} & $\ev$ & 2065 &  251 & 80.36 & 220 & 260 & 150 & 370 \\
\CDF & \cite{CDF-90} & $\ev$ & 1130 &  N/A & 79.91 & 350 & 190 & 240 & 465 \\
     &               & $\uv$ &  592 &  N/A & 79.90 & 530 &  80 & 315 & 620 \\
\CDF & \cite{CDF-95} & $\ev$ & 5718 &  N/A & 80.49 & 145 & 120 & 130 & 230 \\ 
     &               & $\uv$ & 3268 &  N/A & 80.31 & 205 &  50 & 120 & 245 \\
\Dz\ & \cite{D0_Ia}  & $\ev$ & 5982 &  366 & 80.35 & 140 & 160 & 145 & 255 \\
\Dz\ & \cite{D0_Ib}  & $\ev$ & 23068 & 2179 & 80.44 &  70 &  70 &  60 & 115 \\ 
\Dz\ & \cite{D0_EC}  & $\ev$ & 11090 & 1687 & 80.69 & 110 & 190 &  75 & 230 \\
\CDF & \cite{CDF-99}
 & $\ev$ & 30100 & 1600 & 80.47 &  65 &  75 &  55 & 115 \\
    &               & $\uv$ & 14700 & 1800 & 80.47 & 100 &  85 &  55 & 145 \\
\hline
\end{tabular}
\end{center}
\end{table}

In combining the results from the three $\pp$-collider experiments, correlations must be accounted for. Since the Monte Carlo models used by the three experiments were tuned independently based on experimental data, the detector models are certainly independent. The uncertainties due to higher order corrections are dominated by independent experimental uncertainties. The constraints on the $W$ $\pT$ spectra are dominated by statistical fluctuations in the respective $Z$ data samples and are therefore uncorrelated as well. Thus, the only significant correlation originates from the common uncertainty in the structure of the proton. Table \ref{tab:combined} summarizes the combined data.

\begin{table}[ht!]
\begin{center}
\caption{Summary of combined measurements of the $W$-boson mass at $\pp$ colliders}
\label{tab:combined}
\begin{tabular}{@{}lrrr@{}}
\hline\hline
          &            & \multicolumn{2}{c}{uncertainty (MeV)} \\
experiment& \Mw\ (GeV) & total  & correlated  \\
\hline
\UAT              & 80.363 & 371 & 85 \\
\CDF              & 80.433 &  79 & 25 \\
\Dz               & 80.482 &  91 &  9 \\
\hline
\end{tabular}
\end{center}
\end{table}

The individual results combine to
\begin{equation} \Mw(\pp) = 80.452\pm0.060\ \mbox{GeV} \end{equation}
with $\chi^2=0.23$. 

%
\section{MEASUREMENTS OF \Mw\ AT LEP}
\label{sec:lep}

\begin{figure}[htb!]
  \centerline{\hbox{
    \epsfig{file=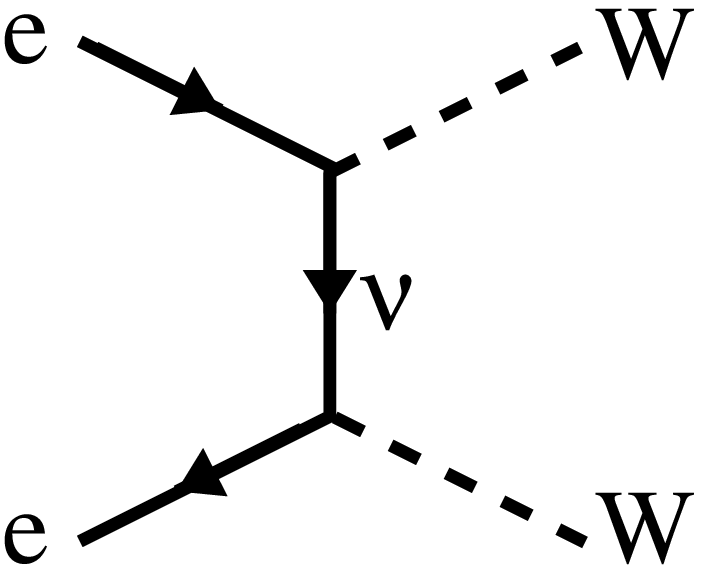,
      width=0.27\textwidth}
    \hspace*{4mm}
    \epsfig{file=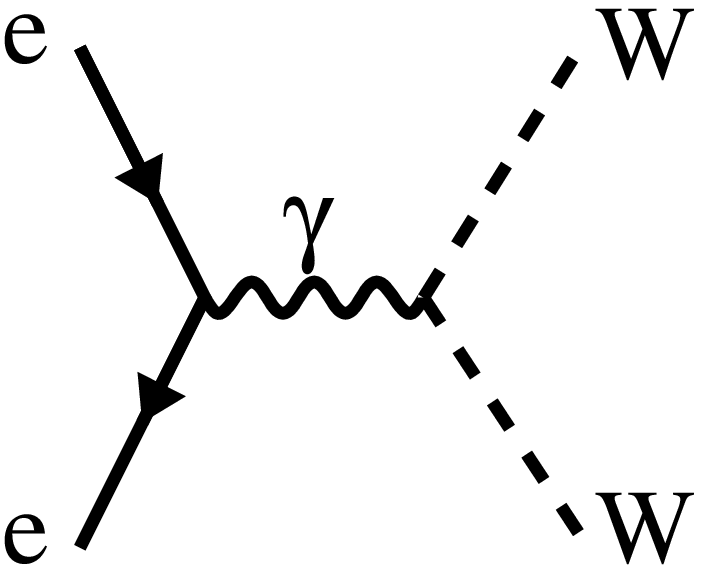,
      width=0.27\textwidth}
    \hspace*{4mm}
    \epsfig{file=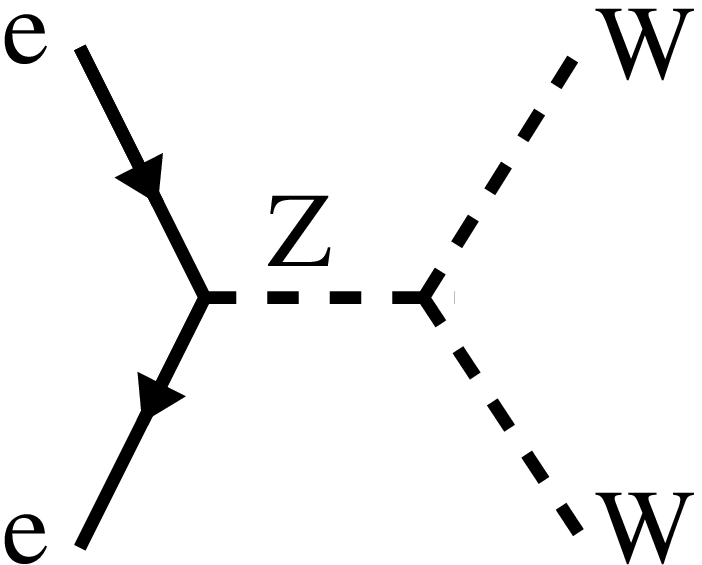,
      width=0.27\textwidth}
    }
  }
  \caption{ The tree level diagrams for the process $\epem\ra\WW$: t-channel 
    neutrino exchange, and s-channel $\gamma$ and $\mrm{Z}$ exchange.}
  \label{fig:lep-cc03}
\end{figure}

From 1989-1995 the Large Electron-Positron collider (LEP) at CERN provided 
$\epem$ collisions at \com\ energies at or near the $Z$-boson mass.
Since 1996, LEP has been running at \com\ energies above the $W$-pair 
production threshold, $\roots\geq 2\Mw$.
The collider provides data to four experiments, \Aleph, \Delphi, \Lt,
and \Opal.  While the LEP1 program afforded precision measurements of the
$Z$-boson mass, the LEP2 program provides the opportunity to precisely measure
the $W$-boson mass\footnote{``LEP1'' refers to data taken from 1989-1995, when
the LEP collider operated at about $\roots=\Mz$, while ``LEP2'' refers to data
taken from 1996-2000 at $\roots=161-205$~GeV.}.
  
At LEP2 energies $W$-bosons are predominantly produced in pairs through the 
reaction $\eeWW$, whose tree level diagrams appear in 
Figure~\ref{fig:lep-cc03}.  Each $W$ subsequently decays either hadronically 
(\qq), or leptonically (\lnu, $\ell = e$, $\mu$, or $\tau$).  There are then
three possible four-fermion final states, hadronic (\WWqqqq), 
semi-leptonic (\WWqqln), and leptonic (\WWlnln), with branching fractions of 
$46\%$, $44\%$, and $10\%$ respectively.  The \WW\ production cross section 
varies from $3.8$~pb at $\roots = 161$~GeV to $17.4$~pb at $\roots = 200$~GeV.
This can be contrasted with the production cross sections for the dominant 
background processes\footnote{Some of the cross sections given here include 
kinematic cuts which restrict the final states to \WW-like parts of 
phase-space.  These cuts are detailed in Reference~\cite{ybkands}.}, 
$\sigma\left(\epem\ra\Zg\ra\qq \right)\sim 100-150$~pb, 
$\sigma\left(\epem\ra\Wenu \right)\sim 0.6$~pb,
$\sigma\left(\epem\ra\Zzee\right)\sim2-3$~pb, and
$\sigma\left(\epem\ra\Zg\Zg\right)\sim0.5-1.5$~pb, where the spread accounts
for variations across the different LEP2 \com\ energies~\cite{ybkands}.  The 
algorithms used to select candidate events exploit the kinematic properties 
unique to the \WW\ final states. The selection algorithms are sensitive to all
possible \WW\ final states and obtain efficiencies of better than about $70\%$
with purities in excess of about $80\%$.

\subsection{Measurement Techniques}
\label{lepintro-meas}

There are two main methods available to measure \Mw\ at LEP2.  The first
exploits the fact that the \WW\ production cross section is particularly
sensitive to \Mw\ at $\roots\approx 2\Mw$. In this threshold region,
assuming SM couplings and production mechanisms, a measure of the 
production cross section yields a measure of \Mw.  In early 1996 each of the 
LEP experiments collected roughly $10\:\mrm{pb}^{-1}$ of data at
$\roots=161$~GeV and determined \Mw\ using the threshold 
technique~\cite{lep161}.

The second method uses the shape of the reconstructed invariant
mass distribution to measure \Mw.  This method is 
particularly useful for $\roots\geq 170$~GeV where the \WW\ production 
cross section is larger and phase-space effects on the reconstructed mass
distribution are smaller.  Each experiment collected roughly 
$10\:\mrm{pb}^{-1}$ at $\roots=172$~GeV~\cite{lep172} in later 1996,
$55\:\mrm{pb}^{-1}$ at $\roots=183$~GeV in 1997~\cite{lep183}, 
$180\:\mrm{pb}^{-1}$ at $\roots=189$~GeV in 1998~\cite{lep189}, and 
$225\:\mrm{pb}^{-1}$ at $\roots=192-202$~GeV in 1999.  Since most of the LEP2 
data have been collected at \com\ energies well above the \WW\ threshold, the 
LEP2 \Mw\ determination is dominated by the direct reconstruction method.
The results reported in this article only include the data taken through the
end of 1998.  

Each method shall be described in greater detail below.

\subsection{Threshold Determination of \Mw}
\label{sec:lepTH}

\begin{figure}[htp!]
  \begin{center}
    \begin{minipage}{4in}
      \epsfxsize = 4in
      \epsffile{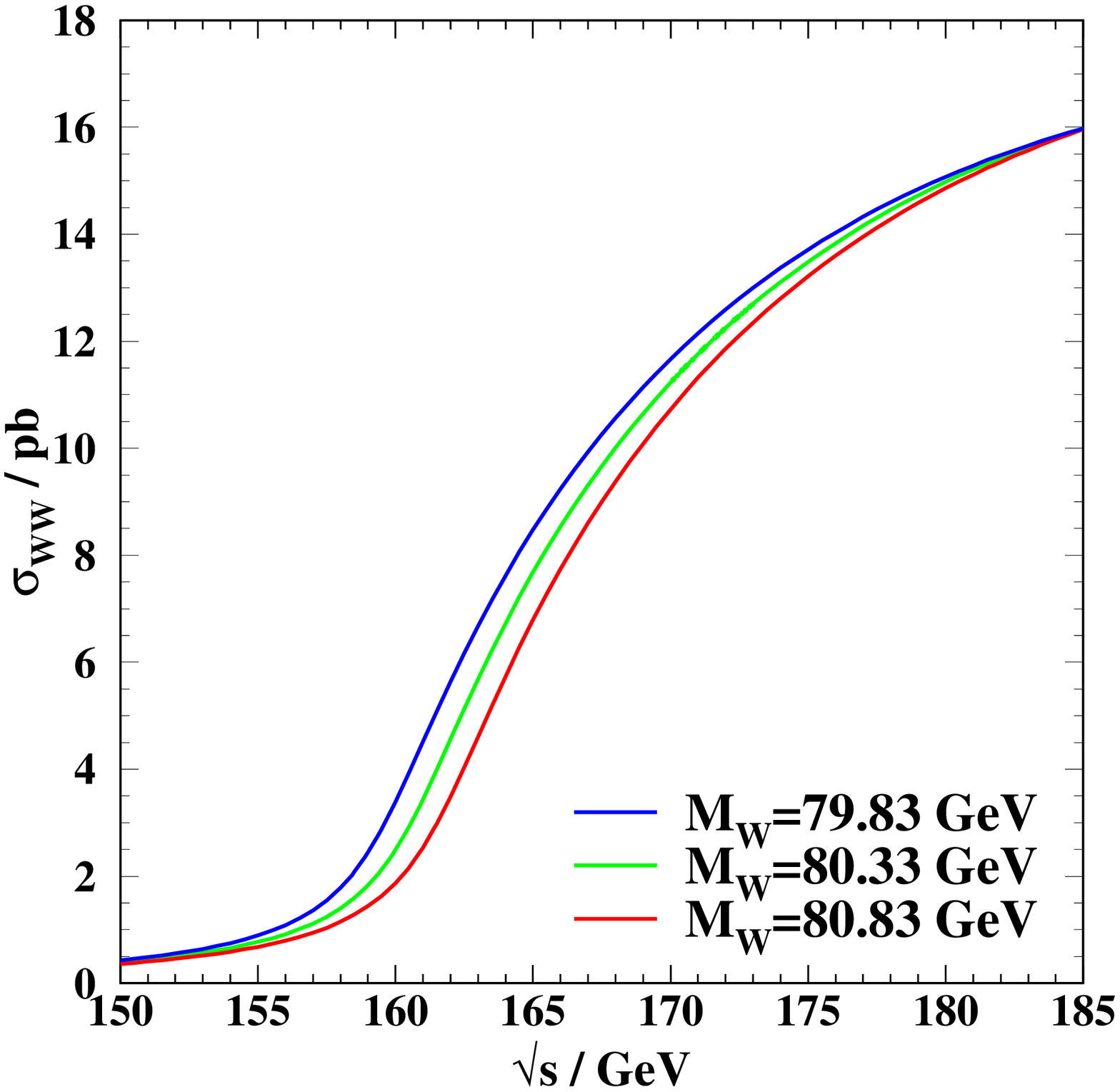}
    \end{minipage}
  \vspace*{-0.15in}
  \end{center}
  \caption{ The $\epem\ra\WW$ cross section as a function of \roots\
   assuming various \Mw\ values.}
  \label{fig:lep-mwvecm}
  \begin{center}
    \begin{minipage}{4in}
      \epsfxsize = 4in
      \epsffile{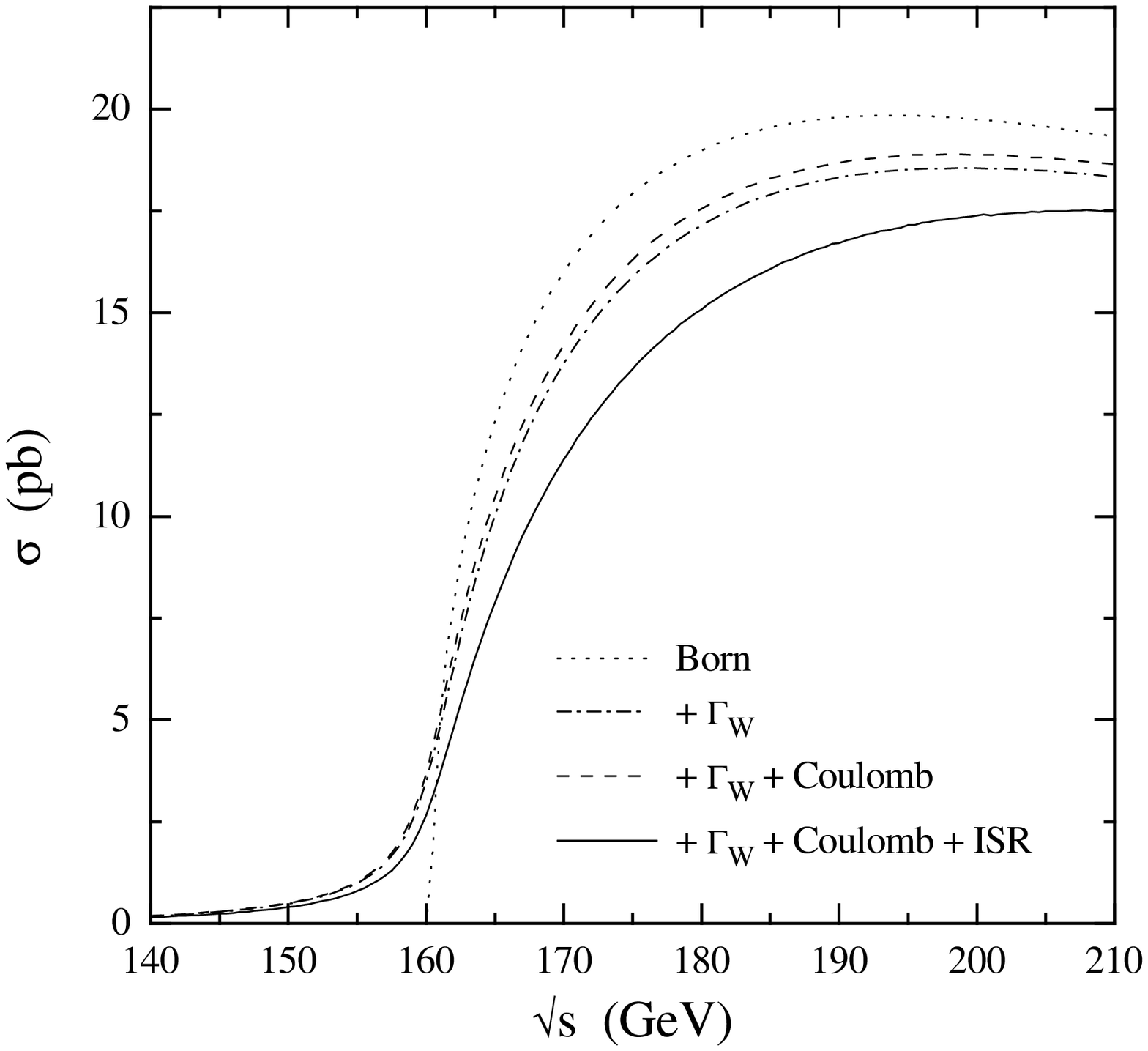}
    \end{minipage}
  \vspace*{-0.15in}
  \end{center}
  \caption{ The $\epem\ra\WW$ cross section as a function of \roots\ using
    the on-shell (Born) approximation, and then including various corrections 
    due to the effect of the $W$-boson width (\Gw), the Coulomb interaction
    between \WW, and ISR\cite{ybkands}.}
  \label{fig:lep-xsww}
\end{figure}

At \com\ energies very near $2\,\Mw$ the \WW\ production cross section, 
$\xsww\equiv\xseeww$, is a strong function of \Mw, so that a measurement of 
\xsww\ can be used to determine the $W$-boson mass.  This is illustrated in 
Figure~\ref{fig:lep-mwvecm}, which plots the $W$-pair production cross section
as a function of \com\ energy for various assumed values of the $W$-boson mass.
Note that for \roots\ significantly above or below  $2\,\Mw$, the various 
curves converge, so that \xsww\ has little sensitivity to \Mw\ at those 
energies.  It is only in the threshold region that the curves significantly 
separate, so that a measure of \xsww\ affords a determination of \Mw.  To 
measure the $W$-boson mass using the threshold method, one needs to 
1) select events, 2) determine \xsww, and 3) extract \Mw\ from the \xsww\ 
determination.  In practice, steps 2) and 3) are not completely independent due
to quantum interference effects, which require that special care be taken when
defining the \WW\ production cross section.  Before discussing each of the 
steps in more detail below, it is useful to detail the \xsww\ calculation.

The $W$-pair production cross section is defined to be the production
cross section for the diagrams given in Figure~\ref{fig:lep-cc03}.  The 
separation between the signal \WW\ production diagrams and background diagrams
resulting in the same four-fermion final states 
(\eg\  $\epem \ra \WW \ra u\ol{d}\ol{u}d$ and
$\epem\ra Z^*Z^* \ra u\ol{u}d\ol{d}$) is complicated by quantum
interference effects.  In addition, the shape of the \xsww\ vs.\ \roots\ curve
is affected by higher order electroweak and QCD corrections.  The effects of 
four-fermion interference and the electroweak and QCD corrections must all be 
sufficiently understood theoretically in order to meaningfully determine \Mw.
 It turns out that the interference effects are small and can be
sufficiently addressed in the background subtraction, as discussed in 
section~\ref{sec:xsmwTH}.  The effects of the higher-order corrections are 
larger and will be further discussed here.

As illustrated in Figure~\ref{fig:lep-xsww}~\cite{ybkands}, the 
\xsww\ vs.\ \roots\ curve gets
smeared out near $2\Mw$ by the effects of the natural width of the $W$-boson, 
\Gw, and by the effects of initial state radiation (ISR).  There are also large
corrections associated with Coulomb interactions between the two $W$s, and some
QCD corrections affecting $\WW\ra\qq\ff$ final states.  The theoretical 
uncertainties associated with calculating the necessary corrections contribute
a $2\%$ uncertainty in \xsww.  These theoretical uncertainties are dominated by
the uncertainty in the Higgs boson mass, which contributes to higher-order 
electroweak loop corrections and is most pronounced near threshold ($1.5\%$).
The remaining uncertainties contribute below the $0.5\%$ level~\cite{ybbandb}. 

\subsubsection{EVENT SELECTIONS}
\label{sec:selectTH}

The statistical uncertainty in \Mw\ determined from the threshold method can be
expressed as
\begin{equation}
  \Delta\Mw(\mrm{stat}) = \sqrt{\xsww} \left|\frac{d\Mw}{d\xsww}\right| 
    \frac{1}{\sqrt{\varepsilon_{\mrm{WW}}\mathcal{LP}}},
\end{equation}
where $\varepsilon_{\mrm{WW}}$ and $\mathcal{P}$ are the \WW\ selection 
efficiency and purity, respectively and $\mathcal{L}$ is the total integrated
luminosity.  From this it is obvious that high efficiency, high purity 
selections are important.  Separate selections are developed for each of the 
main four-fermion final states --- the fully hadronic, the semi-leptonic, and 
the fully leptonic.  Each will be discussed separately.  The algorithms 
employed are quite involved and vary in the details of their implementation 
across the LEP experiments.  In the descriptions below an effort is made to 
simply emphasize the most important discriminating variables and the dominant 
systematic uncertainties.  For detailed descriptions of the selection 
algorithms, the reader is referred to References~\cite{lep161}. 

\subsubsection{\WWqqqq\ EVENT SELECTION}
\label{sec:wwqqqqselTH}

The fully hadronic selection is designed to efficiently select \WWqqqq\ events,
which are characterized by four (or more) energetic hadronic jets, with little 
missing energy or momentum.  The dominant background is from the QCD processes
$\epem\ra\Zg\ra\qq(+n\mrm{g})$, which radiate little energy to ISR.  
Discrimination relies primarily on the fact that the jets in signal events tend
to be higher energy and more spherically distributed than those in background
events.  In addition, to further reduce the QCD background, a kinematic fit can
be employed which requires the two di-jet masses to be approximately equal.
The selections usually require high multiplicity, full energy
events and exploit the unique \WWqqqq\ kinematics  in a multivariate 
discriminant (\eg\ a neural net output) to separate signal from background.  
The typical selection efficiency is about $55\%$ with $80\%$ purity.

For the background estimate, the dominant systematic uncertainty (5\%) is 
associated with modeling the dominant QCD background, estimated by 
comparing data to Monte Carlo using high statistics samples of 
$\epem\ra\Zz\ra\qq$ events from LEP1, and by comparing the estimates from 
various Monte Carlo generators (\ie\ \Pythia\ and \Herwig).  The uncertainty 
associated with the signal efficiency is dominated by comparisons of different
generators (\Pythia, \Herwig, \Ariadne, \Koralw\ and \Excalibur).  
Contributions from uncertainties in the LEP beam energy, \Ebm, and ISR are 
negligible ($<1\%$).  The selection efficiency is also negligibly dependent on
\Mw\ and on the details of modeling color-reconnection and Bose-Einstein 
correlation effects, which are discussed in more detail in 
Section~\ref{sec:lep-syserrdr}. 

\subsubsection{\WWqqln\ EVENT SELECTION}
\label{sec:wwqqlnselTH}

The semi-leptonic selection is designed to efficiently select \WWqqln\ events
and is typically broken into three separate selections, one for each lepton
flavor.  

The \WWqqen\ and \WWqqmn\ events are characterized by two energetic
hadronic jets, a high energy, isolated lepton and large missing momentum 
associated with the prompt neutrino from the leptonically decaying $W$.
The dominant background is from radiative $\epem\ra\Zqq$ events in 
which a hadron or initial state photon is misidentified as a lepton.  
Other background sources include $\epem\ra\Wenu$, $\epem\ra ZZ$
and $\epem\ra Z\epem$ events.  The selections require two hadronic jets,
an identified, energetic (\eg\ $E>25$~GeV), isolated electron or muon
and large missing momentum.  The backgrounds from radiative $\epem\ra\qq$ and
$\epem\ra\Wenu$ events tend to produce missing momentum along the beam axis.  
Requiring a significant missing transverse momentum dramatically reduces these
backgrounds.  The typical selection efficiency is about $70-80\%$ with purities
of around $95\%$.  These selections also select about $5\%$ of \WWqqtn\ events 
due to the leptonic decays of the tau.

The dominant systematic uncertainty associated with the selection 
efficiencies ($2\%$) is due to uncertainties in the Monte Carlo modeling of 
the data and from comparing different Monte Carlo generators for \eeWW\ events.
The dominant systematic uncertainty ($30-50\%$) associated with the 
background estimate is again due to the modeling of the dominant 
$\epem\ra\Zqq$ background and from comparisons of different Monte Carlo 
generators.  For the \WWqqen\ channel, uncertainties from the modeling of 
four-fermion interference, particularly from $\epem\ra\Wenu$ events, estimated 
by comparing the results of different Monte Carlo generators, can increase the 
total background uncertainty to $100\%$.  Since the selections are so pure, 
these relatively large uncertainties in the accepted background cross sections 
translate into very small uncertainties in \xsww.

The \WWqqtn\ events are characterized by two hadronic jets, a $\tau$-decay
jet, and missing momentum associated with two or more neutrinos.  The 
dominant background arises from radiative $\epem\ra\Zqq$ events where a
third jet, often due to soft gluon emission, is misidentified as a 
$\tau$-jet.  These selections are very similar to the \WWqqen\ and
\WWqqmn\ selections except that they identify the $\tau$ as a low-mass, low
multiplicity (1- or 3-prong), isolated jet.  Since the lepton identification
is looser than for the other \WWqqln\ selections, the background tends to
be higher.  Selection efficiencies vary widely among the LEP experiments ---
from $35-45\%$ exclusive of those \WWqqtn\ events identified by one 
of the other selections.  Due to the looser lepton identification 
requirements these algorithms typically select an additional
$3-5\%$ of \WWqqen\ and \WWqqmn\ events failing the above selections. 
The typical purity of this selection also varies widely among the LEP 
experiments, $65-85\%$.

The dominant systematic uncertainty associated with the selection efficiency
($2.5\%$) is due to the modeling of the lepton identification variables, 
estimated by comparing LEP1 data and Monte Carlo, and from the comparison of 
various Monte Carlo generators.  The dominant systematic uncertainty associated
with the estimate of the accepted background cross section ($20\%$) comes from
the modeling of hadron mis-identification, estimated by comparing the data and
Monte Carlo fake rates in LEP1 $\epem\ra\Zz\ra\qq$ events.

\subsubsection{\WWlnln\ EVENT SELECTION}
\label{sec:wwlnlnselTH}

The fully leptonic channel, \WWlnln, is characterized by two high energy, 
isolated, acoplanar leptons.  The selections typically start by requiring a low
multiplicity and large missing transverse momentum.  There are six distinct 
$\ell\ell^{\prime}$ final states ($ee$, $e\mu$, $e\tau$, $\mu\mu$, $\mu\tau$ 
and $\tau\tau$), which have differing dominant background sources.  Potential 
background sources are two photon, $\epem\ra\Wenu$, $\epem\ra\Zzee$, and 
radiative $\epem\ra\Zz\ra\lplm$ events.  In general the \WWlnln\ selection 
involves several independent and overlapping sets of cuts which employ various
combinations of specific electron, muon, and tau identification algorithms.
Backgrounds are usually rejected by requiring large missing energy, large 
transverse momentum and a large lepton-lepton acoplanarity.  The efficiency 
varies widely across the LEP experiments, from about $45\%$ for \Delphi\ and 
\Lt, to about $65\%$ for \Opal\ and \Aleph.  The selection purities are around
$90-95\%$.

The dominant systematic uncertainty associated with estimating the selection
efficiency ($2\%$) is due to the modeling of lepton identification 
variables, specifically those sensitive to FSR modeling (\eg\ isolation 
variables). Comparisons from different Monte Carlo generators also contribute.
The dominant systematic uncertainty associated with estimating the accepted 
background cross sections ($70\%$) arises from limited Monte Carlo statistics 
and comparisons of different Monte Carlo generators.  The effect of detector 
mis-modeling is small owing to the experience gained at LEP1 using 
$\epem\ra\Zz\ra\lplm$ data.  However, veto cuts employed in these low 
multiplicity final states are particularly sensitive to beam related 
backgrounds which are not included in the Monte Carlo. These are estimated 
using random trigger crossings and have the consequence of reducing both the 
signal efficiency and accepted background by a factor of order $0.95-1.0$ with
a relative uncertainty of a few percent.

\subsubsection{DETERMINING \xsww\ AND \Mw}
\label{sec:xsmwTH}

A maximum likelihood procedure is used to determine \xsww.  The likelihood
is usually taken to be the product of Poisson probabilities of observing $N_i$
events when expecting 
$\mu_{i}(\sigma_{WW}) 
  = {\mathcal{L}}\cdot[\sigma_{WW}\cdot{\mathcal{B}}_{i}\cdot\varepsilon_{i}  
   + \sigma_{{\mrm{bgd}}_i}^{\mrm{acc}}]$ events,
where $\varepsilon_{i}$, ${\mathcal{B}}_{i}$, and 
$\sigma_{{\mrm{bgd}}_i}^{\mrm{acc}}$
are the selection efficiency, branching ratio, and accepted background 
cross section, respectively, for the $i$th selection, and $\mathcal{L}$ is the
integrated luminosity.  
In the likelihood calculation, correlations between the channels are properly 
accounted for and Standard Model branching ratios are assumed.  The accepted 
background cross section is assumed independent of \Mw.  Four-fermion 
interference effects are typically accounted for either by applying a 
correction factor, $f_i$, to the product 
$\sigma_{WW}{\mathcal{B}}_{i}\ra\sigma_{WW}{\mathcal{B}}_{i}f_{i}$, or by
adding a factor, $F_i$, to 
$\sigma_{{\mrm{bgd}}_i}^{\mrm{acc}}\ra\sigma_{{\mrm{bgd}}_i}^{\mrm{acc}}
+F_{i}$.
These corrections are negligible ($\left|1-f_{i}\right|<1\%$) for all channels
except the $\qq\enu$ and $\lnu\lnu$ channels which have 
$f_{\qq\enu}\approx 1.09$
($F_{\qq\enu}\approx -0.05$~pb) and $f_{\lnu\lnu}\approx0.97$ 
($F_{\lnu\lnu}\approx +0.01$~pb).  These correction factors are determined by 
comparing the predicted total accepted cross section
(\ie\ signal plus background) calculated including and excluding these 
interference effects.  A systematic uncertainty is estimated by comparing the
predictions from different Monte Carlo generators which include the 
four-fermion (4f) interference effects.
Given the large statistical uncertainty on \xsww\ these corrections do not 
significantly affect the result.  Strictly speaking, the correct manner in
which to account for these interference effects, is to make the replacement 
$\sigma_{WW}\ra\sigma_{\mrm{4f}}$ in the likelihood calculation, where
$\sigma_{\mrm{4f}}$ is the \Mw-dependent 4-fermion production cross section,
including the interference effects.

The measured \WW\ production cross section is then compared to SM predictions 
for \xsww\ dependent on \Mw\ and \Ebm.  The likelihood equation is modified so 
that $\sigma_{WW}\ra\sigma_{WW}(\Mw,\Ebm)$.  The dependence of \xsww\ 
on \Mw\ is calculated using the semi-analytic program \Gentle~\cite{gentle},
which includes the higher order electroweak and QCD corrections.  The results 
from each of the LEP experiments are given in Table~\ref{tab:mw_th}.

In addition to the systematic uncertainties associated with the selection
efficiencies and accepted background cross sections described above, there
are uncertainties due to higher-order corrections affecting 
$\sigma_{WW}(\Mw,\Ebm)$ ($2\%$ at $\roots=161$~GeV as discussed above) and to 
the precision of the LEP determination of \Ebm\ ($\pm27$~MeV at 
$\roots=161$~GeV~\cite{lep_ebm}).

\begin{table}[htb!]
  \caption{ $W$-pair production cross section, \xsww, and $W$-boson mass, \Mw, 
    results for data taken at $\roots=161$~GeV }
  \begin{center}
    \begin{tabular}{|c||c|c|c|} \hline\hline
      exp &$\xsww\pm(\mrm{sta})\pm(\mrm{sys})$ (pb)
          &$\Mw\pm(\mrm{sta})\pm(\mrm{sys})$ (GeV) 
          & $\mathcal{L}\:(\mrm{pb}^{-1})$      \\ \hline
   \Aleph & $4.23 \pm 0.73 \pm 0.19$
          & $80.14\pm0.34\pm0.09$ 
          & $11$                                \\
  \Delphi & $3.67^{+0.97}_{-0.85} \pm 0.19$
          & $80.40\pm0.44\pm0.09$ 
          & $10$                                \\
      \Lt & $2.89^{+0.81}_{-0.70} \pm 0.14$
          & $80.80^{+0.47+0.09}_{-0.41-0.08}$ 
          & $11$                                \\
    \Opal & $3.62^{+0.93}_{-0.82} \pm 0.16$
          & $80.40^{+0.44}_{-0.41}\pm0.10$ 
          & $10$                                \\ \hline
    \end{tabular}
  \end{center}
  \label{tab:mw_th}
\end{table}

\subsubsection{COMBINATION OF \Mw\ DETERMINATIONS FROM THRESHOLD}
\label{sec:comboTH}

\begin{figure}[htb!]
  \begin{center}
    \begin{minipage}{4in}
      \epsfxsize = 4in
       \epsffile{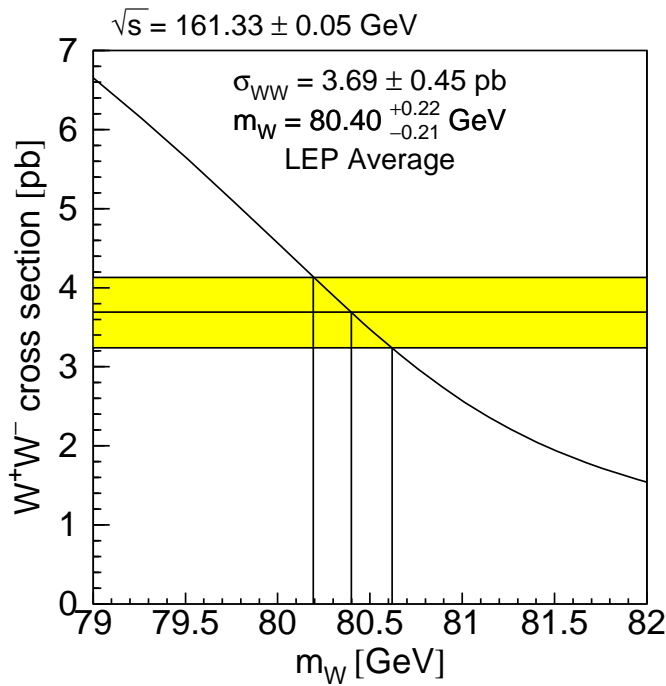}
    \end{minipage}
  \end{center}
  \caption{ The LEP combined \xsww\ (shaded band) is compared to a 
    semi-analytic calculation
    of $\xsww(\Mw,\roots)$ using the LEP average \com\ energy to extract a
    LEP combined determination of \Mw\ using the threshold 
    method\cite{lepmwth}.}
  \label{fig:lep-mwTH}
\end{figure}

The combined determination of \Mw\ is extracted from the LEP combined
measurement of \xsww, which is then compared to the \Gentle\ prediction 
for $\xsww(\Mw,\Ebm)$ assuming SM couplings and the LEP average 
\com\ energy as shown in Figure~\ref{fig:lep-mwTH}.  When making the 
combination the {\it expected} statistical uncertainty is used to calculate 
the weight of each experimental 
measurement so as to avoid biasing the result.  Since the
statistical uncertainties dominate each individual measurement as well as
the combined result, the smallest quoted systematic uncertainty ($0.14$~pb) is
conservatively taken to be fully correlated between experiments.  Note that 
since an individual experiment's weight in the combination is driven by its 
statistical uncertainty, this procedure does not affect the central value of 
the combination and yields a conservative estimate of the combined systematic 
uncertainty.  Combining the four LEP experiments' determinations of \xsww\ 
yields
\begin{equation}
  \xsww = 3.69 \pm 0.45 \:\mrm{pb}
\end{equation}
with a $\chi^2$ per degree of freedom of $1.3/3$.  Using the LEP average \com\ 
energy of $161.33\pm0.05$~GeV and the \Gentle\ prediction, the $W$-boson mass 
is then determined from this threshold (TH) method to be~\cite{lepmwth}
\begin{equation}
  \Mw(\mrm{TH}) = 80.400\pm0.220 (\mrm{exp}) \pm 0.025 (\Ebm)\:\mrm{GeV}.
\end{equation}
The statistical uncertainty dominates the experimental uncertainty,
which has a contribution of approximately $70$~MeV from correlated systematics.
If this method were to be employed in the future (\eg\ at an NLC), a 
potentially limiting uncertainty is due to the modeling of fragmentation
and hadronization, which has a large effect on the \WWqqqq\ channel and is
correlated among the experiments.  This uncertainty is presently the single
largest contribution to the total uncertainty assigned to the LEP combined 
\xsww\ at higher center-of-mass energies (where the statistical uncertainties
are smaller) and contributes approximately a $50$~MeV uncertainty to 
$\Mw(\mrm{TH})$.

\subsection{Direct Reconstruction of \Mw}
\label{sec:lepDR}

As demonstrated in Figure~\ref{fig:lep-mwvecm}, at \com\ energies above 
$170$~GeV the \WW\ production cross section becomes significantly less 
sensitive to \Mw. At these energies one can extract a measure of \Mw\ from the
invariant mass distribution of the $W$-decay products.  The sensitivity to 
uncertainties associated with the modeling of events near the phase-space 
limit ($M_{W^+} = M_{W^-} = \Ebm$) is greatly reduced since 
$(\roots - 2\Mw) \gg \Gw$.  However, as discussed in 
Section~\ref{sec:lep-syserrdr}, the modeling of various final state 
interactions becomes more important in the \WWqqqq\ channel.  To measure \Mw\ 
using this direct reconstruction technique one must 1) select events, 
2) obtain the reconstructed invariant mass, \mrec, for each event, and 
3) extract a measure of \Mw\ from the \mrec\ distribution.
Each of these steps is discussed in detail in the sections below.  

\subsubsection{EVENT SELECTION}
\label{sec:selectDR}

The expected statistical uncertainty on \Mw\ determined from direct 
reconstruction 
will vary inversely with the selection efficiency and purity.  At higher \com\
energies the \WW\ production cross section increases by over a factor of four,
while the dominant background cross sections increase less rapidly, or even 
decrease.  This affords selections with greater efficiencies for the same 
purities relative to those employed for the selection at $\roots=161$~GeV. 
Nevertheless, the algorithms employed are very similar to those described in 
Section~\ref{sec:selectTH} and so will not be further discussed here.  
Typical selection efficiencies (purities) are $85\%$ ($80\%$) for the \WWqqqq\
channel, $90\%$ ($95\%$) for the \WWqqen\ and \WWqqmn\ channels, and $65\%$ 
($85\%$) for the \WWqqtn\ channel.  The \WWlnln\ channel does not significantly
contribute to the determination of \Mw\ from direct reconstruction and is not 
discussed further.  For the high energy data taken through 1999, these 
efficiencies and purities yield approximately $7000$ $\WW\ra\qq\ff$ 
candidate events, about $1100$ of which are non-\WW\ background.  The selection
efficiencies have a total uncertainty of about $1\%$ (absolute) and have a 
negligible effect ($<1$~MeV) on the \Mw\ determination.  The accepted 
background cross sections have a total uncertainty of $10-20\%$ (relative) and
affect the \Mw\ determination at the $10-15$~MeV level 
(cf. Section~\ref{sec:lep-syserrdr}). 

\subsubsection{INVARIANT MASS RECONSTRUCTION}
\label{sec:lepmrec}

For each selected event, an invariant mass is reconstructed from the $W$ decay
products.  There are several methods available for reconstructing the invariant
mass of a $W$ candidate.  The best resolution is obtained by using a 
kinematic fit which exploits the fact that the \com\ energy of the
collision is known {\it a priori}~\footnote{Strictly speaking, this is not true
since any ISR reduces the collision energy to less than twice the beam energy.
The kinematic fits assume no ISR.  The effect of ISR uncertainties is 
incorporated in the total systematic uncertainty discussed in 
Section~\ref{sec:lep-syserrdr} and is small.}. Since the type of fit employed
varies for each final state, each will be discussed separately.  While the
details of the fits differ among the LEP experiments, the important features
are similar.

Selected \WWqqqq\ events are forced into a four-jet configuration using, for 
example, the Durham algorithm~\cite{durham}.  A kinematic fit is then performed
to estimate the reconstructed invariant mass of the $W$ candidates.  A fit 
which incorporates the constraints of energy and momentum conservation (4C fit)
yields two reconstructed invariant masses per event ($\mreca$,$\mrecb$), one
for each $W$-boson in the final state.  A fifth constraint can be incorporated
by neglecting the finite $W$ width and constraining the invariant masses to be 
equal, $\mreca=\mrecb$.  For each event, this 5C fit yields a single 
reconstructed mass, $\mrec$, its uncertainty, $\sigma_{\mrm{rec}}$, and a
fit $\chi^2$-probability.  The fit requires as input the jet momenta, energy, 
and their associated uncertainties.  A complication of the \qq\qq\ final state
is due to the pairing ambiguity --- there exist three possible
jet-jet pairings for a four-jet final state.  This pairing ambiguity gives rise
to a combinatoric background unique to this channel.  Each LEP experiment 
employs a different technique for differentiating among the combinations. 
Typically an experiment will use the best one or two combinations.  The correct
combination is among those used in about $85-90\%$ of the events.  For events 
with the correct pairing, the kinematic fit has a resolution of about $0.7$~GeV
per event and is dominated by the jet angular resolution.  The wrong 
combinations are treated as a background.  It should be noted that the shape of
the combinatoric background is fairly flat (cf. Figure~\ref{fig:l3mw4q}). 
Because of this the \Mw\ determination is not critically dependent on how well
known the fraction of correct pairings is.

Selected \qq\enu\ and \qq\mnu\ events are forced, after removing the lepton 
candidate, into a two jet configuration.  All four LEP experiments use a 
kinematic fit employing energy and momentum conservation constraints and the
equal mass constraint.  Since the prompt neutrino from the 
leptonically decaying $W$ takes three degrees of freedom, this is a 2C 
fit yielding a single reconstructed mass, uncertainty and fit 
$\chi^2$-probability per event.  The fit requires as input the jet and lepton
energy and momenta and their associated uncertainty.  The \qq\enu\ and 
\qq\mnu\ events have a resolution of roughly $1.0$~GeV and $1.1$~GeV, 
respectively, per event.  This resolution is dominated by the uncertainty in
the lepton energy.

Selected \qq\tnu\ events are forced, after removing tracks and clusters 
associated with the $\tau$-decay, into a two jet configuration.  The treatment
of \qq\tnu\ events varies among the LEP experiments.  Basically the invariant
mass of the hadronic system is used, the resolution of which can be improved
by requiring energy and momentum conservation and employing the equal mass
constraint.  The resolution of the \qq\tnu\ events is approximately $1.5$~GeV 
per event and is dominated by the resolution of the jet energies.

\subsubsection{EXTRACTING \Mw }
\label{sec:extmw}

The ensemble of selected events yields a \mrec\ distribution from which a 
measure of \Mw\ is extracted.  There are several methods available for 
extracting \Mw .  \Aleph, \Lt, and \Opal\ all employ a traditional maximum 
likelihood comparison of data to Monte Carlo spectra corresponding to
various \Mw.  In addition to its simplicity, this method has the advantage that
all biases (\eg\ from resolution, ISR, selection, etc.) are implicitly 
included in the Monte Carlo spectra.  The disadvantage of this method is that 
it may not make optimal use of all available information. \Delphi\ employs a 
convolution technique, which makes use of all available information;
in particular, events with large fit-errors are de-weighted relative to fits 
with small fit-errors.  The convolution has the limitations that 
it requires various approximations (\eg\ the resolution is often assumed to be
Gaussian) and often requires an {\it a posteriori} correction as the fit
procedure does not account for all biases, notably from ISR and selection.
All experiments employ an analytic fit of a relativistic Breit-Wigner 
(with $s$-dependent width)+background to 
the data, which also requires {\it a posteriori} corrections, as a cross-check.
Since their dominant systematic uncertainties differ, \Mw\ is  measured from 
the \qq\qq\ and the \qq\lnu\ samples separately.  These are then
combined, taking into account correlations, to yield a final measurement of 
\Mw.  In the results given here, the Standard Model relation between \Mw\ and 
\Gw\ has been assumed~\cite{ybbandb}.

\begin{table}[htb!]
  \caption{LEP Results for the \qq\lnu\ channel for data taken at 
    $\roots=172-189$~GeV}
  \begin{center}
    \begin{tabular}{|cc|} \hline\hline
      exp &$\Mw\pm(\mrm{stat})\pm(\mrm{syst})$/GeV  \\ \hline
      \Aleph  & $80.343\pm0.089\pm0.040$  \\
      \Delphi & $80.297\pm0.141\pm0.064$  \\
      \Lt     & $80.224\pm0.117\pm0.067$  \\
      \Opal   & $80.362\pm0.090\pm0.053$  \\
      LEP     & $80.313\pm0.052\pm0.036$   \\ \hline
    \end{tabular}
  \end{center}
  \label{tab:lep-qqln}
\end{table}

\begin{figure}[htp!]
  \begin{center}
    \psfig{figure=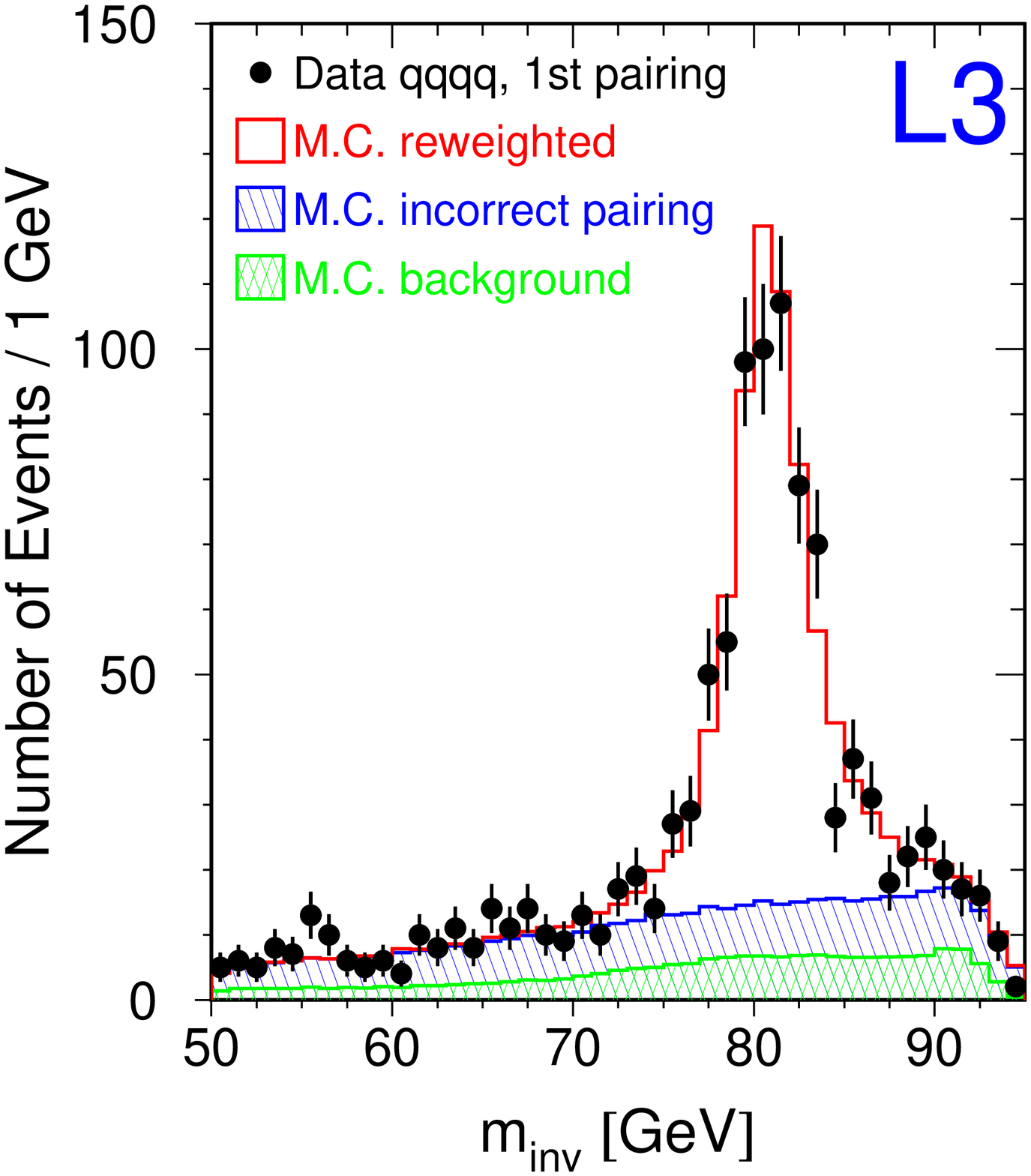,width=3.75in}
  \end{center}
  \vspace*{-0.20in}
  \caption{ The preliminary \Lt\ fit result for the \WWqqqq\ channel using data
    taken at $\roots=189$~GeV.}
  \label{fig:l3mw4q}
%
  \begin{center}
    \psfig{figure=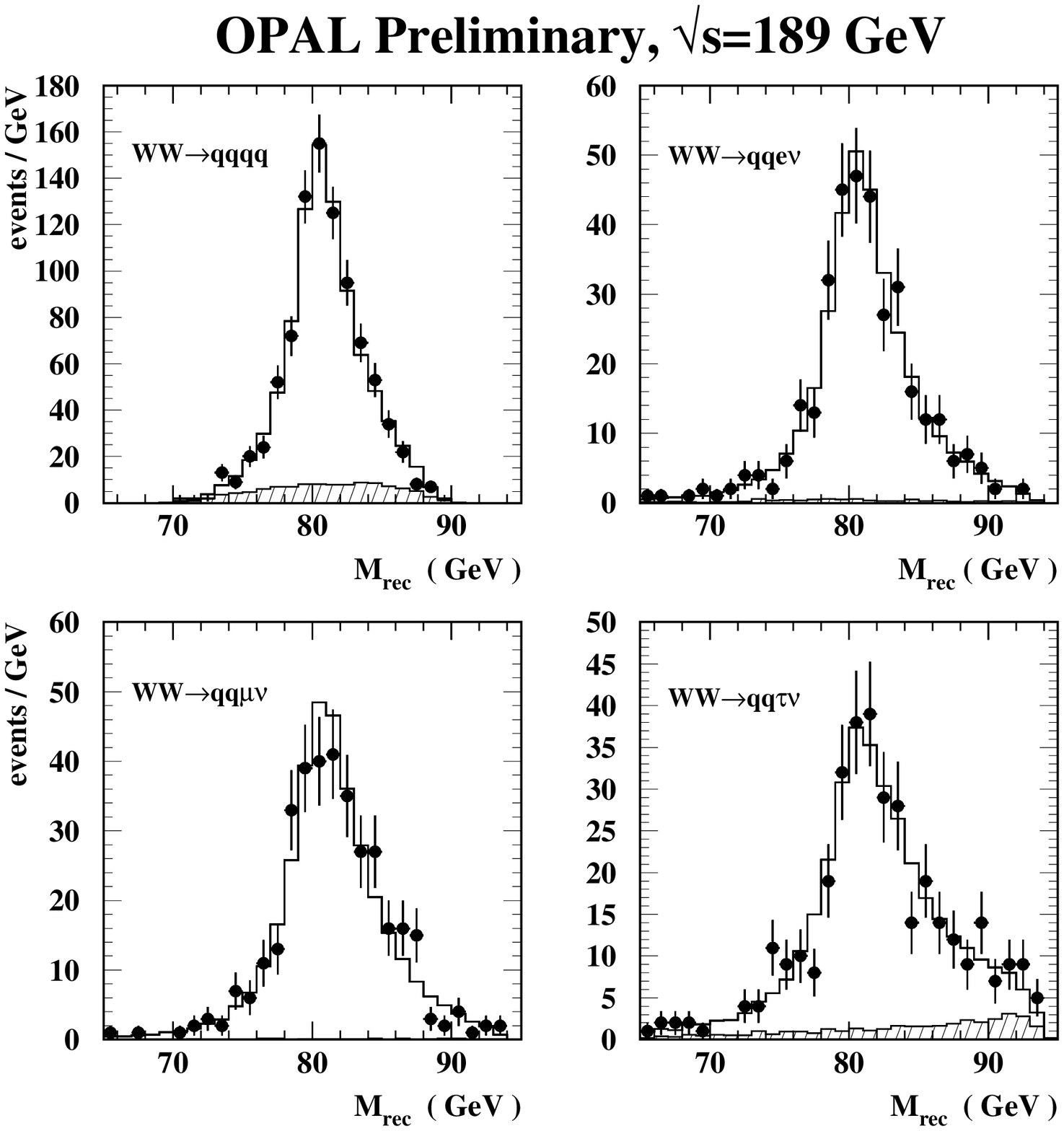,width=3.75in}
  \end{center}
  \vspace*{-0.20in}
  \caption{ Preliminary \Opal\ fit results using data collected at 
    $\roots=189$~GeV.  The points are the data and the open histogram is the 
    fit result.  The non-WW background contribution, as estimated from Monte
    Carlo, is shown as the hatched histogram.}
  \label{fig:mwopal189}
\end{figure}

\begin{table}[htb!]
  \caption{LEP Results for the \qq\qq\ channel for data taken at 
    $\roots=172-189$~GeV}
  \begin{center}
    \begin{tabular}{|cc|} \hline\hline
      exp &$\Mw\pm(\mrm{stat})\pm(\mrm{syst})\pm(\mrm{CR/BE})$/GeV  \\ \hline
      \Aleph  & $80.561\pm0.095\pm0.050\pm0.056$  \\
      \Delphi & $80.367\pm0.094\pm0.037\pm0.054$  \\
      \Lt     & $80.656\pm0.104\pm0.071\pm0.092$  \\
      \Opal   & $80.345\pm0.098\pm0.074\pm0.055$  \\
      LEP     & $80.429\pm0.049\pm0.046\pm0.058$  \\ \hline
    \end{tabular}
  \end{center}
  \label{tab:lep-qqqq}
\end{table}
The results from each LEP experiment, using data collected at 
$\roots=172-189$~GeV~\cite{lep172,lep183,lep189}, are given in 
Table~\ref{tab:lep-qqln} for the \qq\lnu\ 
channel and in Table~\ref{tab:lep-qqqq} for the \qq\qq\ channel\footnote{These
results are based in part on preliminary numbers for the data taken at
$\roots=189$~GeV.}.  Also included is the mass obtained when combining all the
measurements.  In the combination correlations are taken into account as 
described in Section~\ref{sec:lep-comboDR}.  Figure~\ref{fig:mwopal189} shows 
the \Opal\ fits for the data taken at $\roots=189$~GeV.

%

\subsubsection{SYSTEMATIC UNCERTAINTIES}
\label{sec:lep-syserrdr}

The systematic uncertainties for a ``typical'' LEP experiment are given in 
Table~\ref{tab:lep-syserrdr}.  This table should be taken as a general 
guide.  The actual numbers vary by as much as $\pm20$~MeV from experiment to 
experiment.  It is still the case that the total uncertainty of a single 
experiment is dominated by the statistical uncertainty.  As this is the case, 
the experiments are at various stages in developing more sophisticated methods
to estimate the limiting systematic uncertainties.  This largely accounts for 
the range of uncertainties and not any inherent detector or methodological 
advantages.  It should be noted that for all four LEP experiments the 
uncertainties associated with ISR, fragmentation, and four-fermion interference
are limited by the statistics of the comparison. Uncertainties
associated with the selection efficiencies and accepted backgrounds are 
included in the line labeled ``fit procedure''. For the \qq\lnu\ channel the 
largest single contribution to the systematic uncertainty is due to detector 
effects (\eg\ energy scales, resolutions, and modeling).  These uncertainties 
are expected to decrease as more data are collected. For the \qq\qq\ channel 
the dominant systematic uncertainty is due to color-reconnection and 
Bose-Einstein correlation effects.  Each source of uncertainty and the methods
for estimating it are briefly described below.

\begin{description}

\item[Initial state radiation:] 
  Uncertainties due to the modeling of ISR are estimated by comparing the
  \mrec\ distributions of Monte Carlo samples which include ISR corrections to 
  differing orders.  The standard Monte Carlo include corrections to 
  next-to-leading-log ${\mathcal{O}}(\alpha^2)$.  The differences are 
  negligible in samples of several million events.  The uncertainty is 
  conservatively taken to be $10$~MeV.  

\item[Four-fermion interference:] 
  The systematic uncertainty associated with the modeling of four-fermion 
  interference effects is usually estimated by comparing fit results using 
  Monte Carlo reweighting techniques which employ matrix
  element calculations including and excluding the interferences.  The 
  differences are small and the comparisons often statistically limited.

\item[Fragmentation:] A variety of methods is employed to estimate the 
  uncertainty associated with fragmentation modeling.  Typically, LEP1 data 
  are used to constrain model parameters.  Those parameters are then varied in 
  several Monte Carlo samples, which are re-fit as data.  The results are 
  compared to a Monte Carlo sample with default parameters.  The differences 
  are typically small
  except for the \WWqqqq\ channel, where they are on the order of $30$~MeV.
  Since this systematic is strongly correlated among the LEP experiments, it
  is one of the dominant systematic uncertainties in the LEP combined 
  measurement.  \Delphi\ employs an alternative method for estimating these
  uncertainties~\cite{lep183}, in which two $\Zz\ra\qq$ events are boosted
  to the appropriate \com\ energy and overlayed in data and Monte Carlo; the 
  two samples are treated as data and their resulting fit masses compared.

\item[Detector effects:] 
  Calibration data collected at the start and end of each year at $\roots=\Mz$,
  are used to establish limits on the uncertainties associated with detector 
  modeling in the Monte Carlo.  Uncertainties in energy scale and resolution 
  are estimated using $\Zz\ra\epem,\mpmm,\qq$ data.  The linearity is checked
  using Bhabha, $\epem\gamma$, and 3-jet events in data collected at higher
  \com\ energies.  The angular resolutions are similarly determined. Typically
  the jet and lepton energy scales are known to within $0.5$\%, while the 
  modeling of the angular and energy resolutions have associated uncertainties
  of the order $5-20$\% depending on polar angle.  These uncertainties are
  propagated to the \Mw\ measurement by comparing the fit results of Monte 
  Carlo samples in which the appropriate quantity has been scaled or varied to
  the results from the default Monte Carlo sample.  The observed differences 
  are used as estimates of the associated uncertainty on \Mw\ due to detector 
  modeling effects.

\item[Fit procedure:] The type of effects considered depend on the fit method 
  employed to extract \Mw\ from the \mrec\ distribution.  These include 
  uncertainties associated with the background normalization and shape, and fit
  biases.  In general the total effect is 
  very similar across methods.  The background normalization is varied within
  uncertainties determined from a dedicated \xsww\ analysis yielding small
  effects.  The shape of the background \mrec\ distribution is cross-checked
  using data where possible and compared across different Monte Carlo 
  generators otherwise, also yielding small effects.  The linearity of the fit
  methods is determined from Monte Carlo samples generated assuming various 
  \Mw\ values.  These samples are also used to verify that the statistical 
  uncertainty is accurate. For the convolution and Breit-Wigner methods, these
  Monte Carlo samples are used to calculate the necessary bias corrections, 
  whose uncertainties are then propagated to the final uncertainty.

\item[Beam energy:] 
  The uncertainty in the beam energy enters via the constraints employed 
  by the kinematic fit and should be of order $d\Mw=\Mw\frac{dE_{bm}}{E_{bm}}$.
  The effect on the measured \Mw\ is estimated by re-fitting all data changing 
  the \Ebm\ used in the fit and calculating the mean difference in fitted 
  \mrec\ on an event-by-event basis.  The beam energy is estimated using the
  method of resonant depolarization~\cite{lep_rdp}; the last energy point for
  which the depolarization method works is $\Ebm\approx60$~GeV.  An 
  extrapolation is required to estimate the beam energies at which the data
  are taken, $\Ebm\approx100$~GeV.  The uncertainty in the beam energy is
  about $20$~MeV and is dominated by the uncertainty in this extrapolation.
  With the addition of more resonant depolarization data and new techniques, 
  it is expected that the uncertainty on \Mw\ due to uncertainties in \Ebm\ 
  will be reduced to around $10$~MeV~\cite{lep_ebm,lep_spec}.  The spread
  in the beam energy, $\sigma_{\Ebm}\approx150$~MeV~\cite{lep_ebm}, has a 
  negligible effect on \Mw.

\item[CR/BE:] 
  Since the typical decay distance of the $W$-bosons, 
  $1/\Gamma_{W}\approx0.1$~fm, is much smaller than the typical fragmentation 
  radius, $1/\lambda_{QCD}\approx 1$~fm, 
  the decay products originating from {\emph{ different}} $W$-bosons cannot be 
  considered as independent --- \ie\ they can ``talk'' to each other.  The 
  modeling of this cross-talk in the Monte Carlo spectra used to extract \Mw\
  is an additional source of systematic uncertainty in the \WWqqqq\ channel.  
  The cross-talk can arise through two mechanisms, Bose-Einstein correlations 
  (BE) and color reconnection effects (CR)~\cite{ybkands}\cite{lep_becr}.  The 
  modeling uncertainty is estimated separately for BE and CR and is
  model dependent in both cases.  In each case Monte Carlo samples employing 
  implementations of various CR/BE models are treated as data and an \Mw\ bias
  is estimated.   The systematic uncertainty is chosen to include the 
  full range of variation among the models explored and is on the order of
  $50-60$~MeV.
\end{description}
It should be noted that there has been recent progress in experimentally 
constraining the available CR models by comparing event shape and charged 
particle multiplicity distributions as predicted by various Monte Carlo models
(both including and excluding CR effects) with those observed in the 
high-energy data.  In addition, studies using LEP1 data can also be used to 
test the available models~\cite{olep1cr}.  On the basis of these studies, some 
of the models have been excluded as they fail to adequately describe the 
data~\cite{o183cr}, thus enabling a reduction in the associated systematic 
uncertainty (from $\approx100$ to $\approx50$~MeV).  For a more complete 
discussion, see Reference~\cite{o183cr}.   Additional data should help to 
further constrain the remaining CR models and thus improve this uncertainty 
further.

\begin{table}[htb!]
  \caption{ Table of systematic uncertainties on \Mw\ from direct 
    reconstruction for a ``typical'' LEP experiment }
  \begin{center}
    \begin{tabular}{lcc} \hline\hline
      systematic & \multicolumn{2}{c}{uncertainty (MeV)} \\
      source     & \qq\lnu & \qq\qq \\ \hline
      initial state radiation &  10 &  10  \\ 
      four-fermion            &  10 &  10  \\
      fragmentation           &  25 &  30  \\
      detector effects        &  30 &  30  \\
      fit procedure           &  20 &  20  \\
      Sub-total               &  46 &  49  \\
      beam energy             &  17 &  17  \\
      CR/BE                   &  -- &  60  \\ \hline
      Total                   &  49 &  79  \\ \hline
    \end{tabular}
  \end{center}
  \label{tab:lep-syserrdr}
\end{table}

\subsubsection{COMBINATION OF \Mw\ DETERMINATIONS FROM DIRECT RECONSTRUCTION}
\label{sec:lep-comboDR}

Each of the LEP experiments provides their measured $W$-boson mass for the
fully-hadronic and semi-leptonic channels separately for each \com\ energy
along with a matrix of associated uncertainties.  The uncertainties are
broken down into four components:
\begin{enumerate}
  \item uncertainties uncorrelated between channels and experiments 
    (\eg\ the statistical uncertainty or background normalization and shape 
    uncertainties)
  \item uncertainties correlated among the channels of a given experiment,
    but uncorrelated between experiments (\eg\ detector modeling 
    uncertainties)
  \item uncertainties uncorrelated between the channels, but correlated
    among the experiments (\ie\ CR/BE uncertainties)
  \item uncertainties correlated between the channels and among the 
    experiments (\eg\ ISR, fragmentation, \Ebm\ uncertainties).
\end{enumerate}
In this way the correlations between channels and among experiments are 
accounted for.  The correlation of the \Ebm\ uncertainty across the different
years is also taken into account.  The results for the combined \qq\qq\ 
and \qq\lnu\ channels are given in the last lines of Tables~\ref{tab:lep-qqln} 
and \ref{tab:lep-qqqq} and are $25\%$ correlated with a $\chi^2/dof = 17.9/20$.
Combining all the direct reconstruction (DR) results into a single mass yields
\begin{displaymath}
  \Mw(\mrm{DR}) = 80.347 \pm 0.036 (\mrm{stat}) \pm 0.036(\mrm{syst})
    \pm 0.020 (\mrm{CR/BE}) \pm 0.017 (\Ebm)\:\mrm{GeV}
\end{displaymath}
where the uncertainties associated with CR/BE modeling and with the LEP
beam energy have been listed separately~\cite{lepmwdr}.  The dominant 
systematic uncertainty is associated with the fragmentation model, which 
is correlated among the experiments (they all employ the same models in their 
Monte Carlo) and contributes an uncertainty of approximately $20$~MeV.  The 
effect of the CR/BE uncertainty is to de-weight the \qq\qq\ measurements 
relative to the measurements in the \qq\lnu\ channels.

\subsection{Combination of LEP Results}
\label{sec:lep-comboMw}

The \Mw\ determination using the threshold method is combined with the 
determination using the direct reconstruction method taking account of the
correlations.  In particular, the systematic uncertainties associated with
the LEP beam energy, and the modeling of ISR, fragmentation, and 
four-fermion interference effects are taken as correlated.  Note that the 
weight of the
threshold determination of \Mw\ in the combination is driven by the statistical
uncertainty of that measurement.  The LEP combined result, 
assuming the Standard Model relation between the $W$ decay width and mass, is 
\begin{equation}
  \Mw(\mrm{LEP}) = 80.350 \pm 0.056 \:\mrm{GeV}, 
\end{equation}
where the uncertainty includes both statistical and systematic uncertainties
and is dominated by the determinations using direct reconstruction 
methods~\cite{lepmwdr}.

%
\section {WHAT DO THESE MEASUREMENTS TELL US?}
\subsection {Combination of Results}

Direct measurements of the $W$-boson mass have been performed in two kinds of
experiments, the production of $W$-bosons in $\pp$ collisions and the
production of $\WW$ pairs in $\epem$ collisions. 

Until 1996, $\pp$ collisions were the only source of $W$-bosons. The advantage
of $\pp$ colliders lies in the large $W$ production cross section and the 
low background levels. The $\pp$ data give about 100{,}000 $\Wlv$ candidate 
events with about 97\% purity. The production of $Z$-bosons, dynamically and
kinematically very similar to $W$-boson production, provides a very
convenient control data sample. The disadvantage of $\pp$ colliders is that
the parton center-of-mass frame is not known on an event-by-event basis and
therefore systematic effects arising from the structure of the protons must
be understood. The combined $\pp$-collider measurement is 
$\Mw(\pp) = 80.452\pm0.060$ GeV.

Since 1996, $\epem$ collisions with enough energy to produce pairs of 
$W$-bosons are available. The advantage of $\epem$ collisions is that since the
initial particles are point like and so the center-of-mass energy of the 
collision is known, kinematic fits can be employed to fully reconstruct events
and thus yield invariant mass resolutions comparable to the $W$-boson width. 
The disadvantage of \epem\ colliders is that the \WW\ production cross section
is two orders of magnitude smaller than at $\pp$ colliders, resulting in 
smaller and less pure event samples (about 22{,}000 events with about 90\% 
purity).  In addition, the modeling of final-state interactions in \WWqqqq\
events must be understood. The combined LEP
measurement is $\Mw(\mrm{LEP}) = 80.350\pm0.056$ GeV.

The two determinations of the $W$-boson mass are completely uncorrelated. 
A combination of both results is simple, resulting in a world average of
\begin{equation} \Mw = 80.398\pm0.041\ \mbox{GeV} \end{equation}
with a $\chi^2$ of $1.6$. Having two independent, precise determinations of this parameter
in agreement with each other lends significant credibility to the results.
 
Within the framework of the Standard Model, the measurement of the $W$-boson 
mass determines the radiative corrections, $\Delta r$, in 
Equation~\ref{eqn:ewksmrc}. These corrections have a large contribution from the running of the electromagnetic coupling. We can absorb this into the value of $\alpha$ by writing
\begin{equation}
{\alpha\over1-\Delta r}={\alpha(\Mz^2)\over1-\Delta r_{ewk}}.
\end{equation}
For the residual contribution from electroweak loops, we find \linebreak
$\Delta r_{ewk} = -0.0268\pm0.0027$, about 10 standard deviations from zero.

\subsection {Comparisons and Constraints within the Standard Model}

The Standard Model provides us with a framework that allows us to relate the
measurements from many processes that involve the electroweak
interaction. The main sources of such information are measurements of the
properties of the $Z$-boson at LEP1 and SLC, the study of deep inelastic
neutrino scattering at Fermilab, and the measurement of the mass of the top
quark at the Tevatron. 

LEP1 and SLC have provided us with a wealth of very precise measurements of
the properties of the $Z$-boson \cite{LEP-EWK}. At tree level, the properties
of the $Z$-boson are determined by its mass, the weak mixing angle,
and the fine structure constant. Radiative corrections are dominated by the
masses of the top quark and the 
Higgs boson. The wonderful success of the Standard Model lies in all 
measurements being consistent with single values of these parameters.  
The mass of the $Z$-boson is measured directly from the line shape, and the
fine structure constant, evolved to $Q^2=\Mz^2$ is derived from measurements
of $R$, the ratio of the $\epem$ cross sections to hadrons and to $\mpmm$. 
The other three parameters are extracted from a fit to the measurements. 
The $W$ mass then follows from Equation~\ref{eqn:ewksmrc}.

The CCFR \cite{CCFR} and NuTeV \cite{NuTeV} experiments at Fermilab measure the
 ratio of charged current and neutral current interactions of neutrinos. This 
ratio depends directly on $1-\Mw^2/\Mz^2$. From the measured value \linebreak
$1-\Mw^2/\Mz^2=0.2255\pm0.0021$\footnote{CCFR and NuTeV combined.}, a value 
for the $W$-boson mass of $\Mw=80.250\pm0.109$ GeV can be derived.

At the loop level many other parameters contribute (mostly negligible)
corrections to the tree level values. Due to the large mass difference
between the top and bottom quarks, radiative corrections involving top-quark
loops are important. The \CDF\ and \Dz\ Collaborations have measured the top
quark mass directly \cite{Mtop}. Their combined value is $\Mt = 174.3\pm5.1$
GeV. 

A fit of the Standard Model to all measurements except the direct
measurements of the $W$-boson mass returns \cite{mwmhindirect} 
$\Mw = 80.381\pm0.026$ GeV as its preferred value. This value is in excellent agreement with the
combined direct measurements, in support of the validity of the Standard Model.
The mass of the Higgs boson is the only parameter which is not measured
experimentally. Loops containing Higgs bosons also contribute important
radiative corrections. A fit to all electroweak data, including the measurements of the $W$-boson mass, prefers $\Mh = 77^{+69}_{-39}$ GeV for the mass of the Higgs boson \cite{mwmhindirect}. 

\begin{figure}[htb!]
  \begin{center}
    \psfig{figure=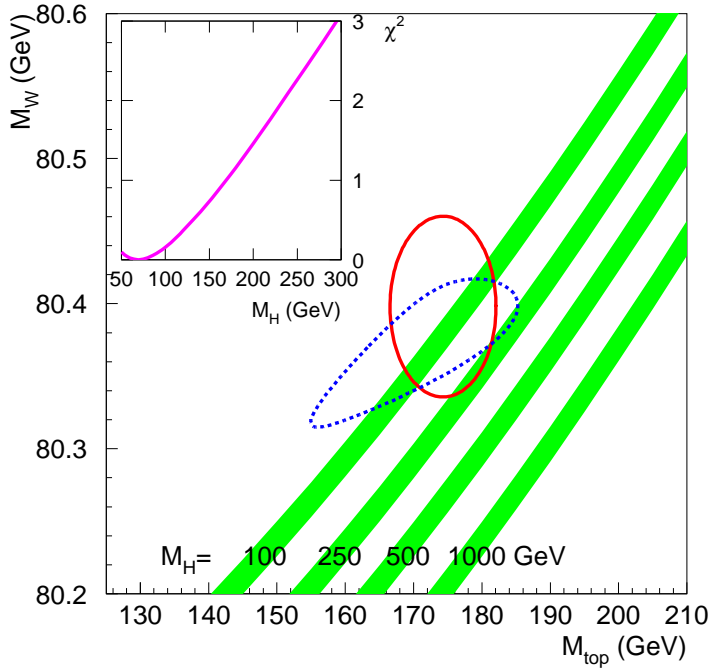,width=4in}
    \caption{Comparison of direct measurements of the $W$-boson and top-quark 
      masses with indirect measurements \cite{mwmhindirect} and predictions of
      the Standard Model~\cite{Degrassi}. The indirect constraint is in part 
      based on preliminary data.}
    \label{fig:MwMt}
  \end{center}
\end{figure}

The Higgs-boson mass can also be constrained based on the measured values of the $W$-boson and top-quark masses alone. This is shown graphically in
Figure~\ref{fig:MwMt}. The shaded bands indicate the values of the $W$-boson mass predicted by the Standard Model as a function of the top quark mass, for given values of the Higgs-boson mass \cite{Degrassi}. The width of the bands indicates the variation due to the uncertainty in $\alpha(\Mz^2)$ \cite{alpha}, which dominates the uncertainty in the predictions. The ellipse indicates the two-dimensional 68\% confidence-level interval defined by the measured mass values. The inset shows a plot of $\chi^2$ between the measured values and the predictions as a function of the Higgs-boson mass. The preferred Higgs-boson mass is $71^{+96}_{-51}$ GeV. Values above 277 GeV are excluded at 90\% confidence level. The dashed contour shows the 68\% confidence-level interval from the fit to all other electroweak data \cite{mwmhindirect}. 

\subsection {Constraints Outside the Standard Model}

Any particle that couples to the $W$-boson can contribute loop corrections to the value of 
the $W$-boson mass. Thus a measurement of the $W$-boson mass does not only test the 
Standard Model, but it is, at least in principle, also sensitive to non-standard physics. 
In the minimal supersymmetric model, corrections could increase the $W$-boson mass by as 
much as 250 MeV \cite{SUSY}. The correction is larger, the lower the scale of 
supersymmetry breaking. If the scale of supersymmetry breaking is more than a few 
hundred GeV, supersymmetry decouples from Standard Model physics and the effects of 
supersymmetric loop corrections on the $W$-boson mass become small. Supersymmetric 
particles that give large corrections must be relatively light and would also be the 
first ones to be seen in direct searches. Thus, precision measurements are unlikely to 
increase the sensitivity of direct searches for supersymmetric extensions of the standard 
model.

%
\section{FUTURE PROSPECTS}

The precision on the world average $W$-mass is expected to improve 
significantly over the next 5 years, and could possibly improve dramatically 
over the next decade or so.

By the end of 2000, the LEP experiments will have more than doubled the 
statistics of their \WW\ data sets relative to what has been included in this
review.  The inclusion of the additional data will yield a statistical 
uncertainty for the combined LEP measurement of \Mw\ of about $25$~MeV.  
Already a significant effort has been made to reduce the systematic 
uncertainties, particularly those associated with the detector energy scales 
and resolutions.  It is expected that these errors can be brought to the 
$20$~MeV level.  The additional constraints afforded by the LEP spectrometer 
project\cite{lep_spec} and by additional depolarization data make it likely 
that the uncertainty associated with the LEP beam energy will be reduced to 
roughly $10$~MeV.  It is difficult 
to predict how the systematic uncertainties associated with the modeling of
Bose-Einstein and color-reconnection effects in the \WWqqqq\ channel will 
evolve.  While it is true that the additional data will provide more 
stringent tests, it is unknown whether the additional sensitivity will actually
reveal a discrepancy large enough to reject any of the remaining viable models.
Assuming none of the remaining models are rejected, so that
the CR/BE uncertainty remains the same, the projected total uncertainty on 
\Mw\ at the end of LEP2 would be of order $35-40$~MeV from the LEP combination.
If the CR/BE uncertainty can be reduced to less than $15$~MeV, it may be 
possible for the LEP combined determination to reach a total uncertainty of 
$30$~MeV.

In spring of 2001, the \CDF\ and \Dz\ experiments will start taking data
at the Fermilab TeVatron.  They anticipate collecting $2-3\:\mrm{fb}^{-1}$ of
data by 2004, which should give a five-fold improvement in the statistical
uncertainty of the TeVatron \Mw\ measurement. The systematic uncertainties 
associated with the energy scale and other detector effects are dominated by
the statistics of the $Z$ control samples and are expected to scale 
accordingly.  On the other hand, the systematic uncertainty associated with 
the $W$-production modeling does not scale directly with statistics and may 
improve only moderately to about $20$~MeV.  The uncertainty from 
the combined TeVatron \Mw\ determination is expected to be about
$30$~MeV\cite{TeV2000}. 

On the time-scale of the next five years, it is expected that the world 
average $W$-mass will have a total uncertainty of $20-25$~MeV --- a factor of 
two improvement relative to the present uncertainty.  The SM constraint on \Mh\
afforded by the \Mw\ measurements alone will be comparable to that afforded by
the $\sin\theta_W$ measurements of LEP and \SLD, which presently yield 
an uncertainty of order $\Delta\Mh = \Mh$.  Although these two sets of 
constraints are correlated through \Mt-dependent corrections, it is still 
interesting to compare them since they have differing sensitivities to the 
various radiative correction terms.  A significant improvement to the SM 
constraints on \Mh\ requires a more precise determination of \Mt.  On the same
time-scale, the TeVatron experiments are expected to measure \Mt\ 
to $2-3$~GeV\cite{TeV2000}.  Including this 
improvement should yield constraints on \Mh\ with uncertainties on the order of
$\Delta\Mh = 0.5\Mh$  (assuming that the fits continue to give a central 
value of $\Mh\sim{\mathcal{O}}(100)$~GeV)\cite{Degrassi}.  

Looking further ahead, on the time-scale of $5$-$10$ years, it is possible that
the LHC experiments, CMS and ATLAS, will measure the $W$-mass to a precision of
$15$~MeV and \Mt\ to $2$~GeV\cite{LHCguess}.  And on the time-scale of 
$>10$ years, future high luminosity \epem\ or \mpmm\ colliders might yield
the statistics to envision a $<10$~MeV measurement of \Mw\ using the threshold
method and a $<1$~GeV measurement of \Mt\cite{NLCguess}\cite{FMCguess}.  If 
achieved, such precision measurements would yield constraints on \Mh\ with 
uncertainties of ${\mathcal{O}}(1-10)$~GeV - which we can hope will offer, by 
that time, a very interesting comparison with the directly measured \Mh!

%
\section{CONCLUSIONS}

The mass of the $W$ boson has been measured by many experiments at $\pp$ and 
$\epem$ colliders. 
All measurements are in good agreement. The world average of all measurements 
of the $W$-boson 
mass is $80.398\pm0.041$ GeV. 
Based on measurements of other parameters, the Standard Model of the 
electroweak interactions 
leads to a prediction of $80.381\pm0.026$ GeV for the mass of the $W$ boson, 
in excellent 
agreement with the measured value. In the framework of the Standard Model this
measurement of 
the $W$-boson mass, together with the measurement of the top-quark mass, 
constrains the 
Higgs-boson mass to values below 280 GeV at 90\% confidence level. Over the 
coming decade, 
a reduction in the uncertainty of the direct measurement of the $W$-boson mass
of at least a factor two is expected. As the top quark mass is measured more 
precisely and the reach of 
searches for the Higgs boson increases, the comparison of the indirect 
constraint on the Higgs-boson mass and its direct measurement or exclusion region will become one of 
the most interesting tests of the Standard Model. This test will for the first
time close in on the symmetry breaking sector of the Standard Model about which
very little is presently known.

\section*{Acknowledgments}

We would like to thank our colleagues at LEP and the Tevatron, with whom we 
had the privilege to work on these exciting measurements. In particular, we 
should like to thank W.~Carithers, J.~Goldstein, A.~Kotwal, M.~Lancaster, M.~Narain, H.~Weerts, J.~Womersley, N.~Watson and D.~Wood for many helpful comments and 
suggestions on the manuscript. This work is partially supported by the U.S. 
Department of Energy under contracts with Fermi National Accelerator 
Laboratory and Boston University.   One of us (UH) also acknowledges the 
support of the Alfred P. Sloan Foundation.

%



\clearpage
\begin{thebibliography}{99}

  \bibitem{ewk} Glashow SL, \Journal{\NPA}{22}{579}{1961}; Weinberg S,
    \Journal{\PRL}{19}{1264}{1967}; Salam A, 1968.  In {\it Elementary Particle
    Theory}, ed. N Svartholm. Stockholm: Almquist and Wiksell. 367 pp.

  \bibitem{Wdiscovery} \uaone, \Journal{\PLB}{122}{103}{1983}; 
    \uatwo, \Journal{\PLB}{122}{476}{1983}.

  \bibitem{SppS} \Journal{\PLB}{107}{306}{1981}.

  \bibitem{UA1} \uaone, \Journal{\ZPC}{47}{11}{1990}.

  \bibitem{UA2} \uatwo, \Journal{\ZPC}{47}{11}{1990}.

  \bibitem{UA2-90} \uatwo, \Journal{\PLB}{241}{150}{1990}.

  \bibitem{CDF} \cdf, \Journal{\NIMA}{271}{387}{1988}.

  \bibitem{D0} \dzero, \Journal{\NIMA}{338}{185}{1994}.

  \bibitem{TeV} Edwards HT, 
   \Journal{\it Annual Reviews of Particle and Nuclear Science}{35}{605}{1985}.

  \bibitem{aleph} \Aleph\ Collaboration, \Journal{NIMA}{294}{121}{1990}.

  \bibitem{delphi} \Delphi\ Collaboration, \Journal{NIMA}{303}{233}{1991};
    \Journal{NIMA}{378}{57}{1996}.

  \bibitem{lthree} \Lt\ Collaboration, \Journal{NIMA}{289}{35}{1990};
    \Journal{NIMA}{349}{345}{1994}; \Journal{NIMA}{351}{300}{1994};
    \Journal{NIMA}{374}{293}{1996}; \Journal{NIMA}{381}{236}{1996};
    \Journal{NIMA}{383}{342}{1996}.

  \bibitem{opal} \Opal\ Collaboration, \Journal{NIMA}{305}{275}{1991};
    Anderson BE \etal, 
    \Journal{IEEE Transactions on Nuclear Science}{41}{845}{1994};
    \Opal\ Collaboration, \Journal{NIMA}{403}{326}{1998}.

  \bibitem{ewktexts} More detailed discussions and derivations can be found,
   for example, in this textbook: Aitchison IJR and Hey AJG 1989. {\it Gauge
   Theories in Particle Physics}, 2nd ed.  Philidelphia: IOP Publishing Ltd.
   550 pp.

  \bibitem{higgs} Higgs PW, \Journal{\PL}{12}{132}{1964}; Higgs PW,
    \Journal{\PRL}{13}{508}{1964}; \Journal{\PRL}{145}{1156}{1966}.

  \bibitem{rcreview} A more detailed discussion and review of electroweak
    radiative corrections can be found, for example, in this article:
    Montagna G, Nicrosini O, Piccinni F, {\it Precision Physics at LEP},
    to appear in {\em Rivista del Nuovo Cimento}, \verb+ hep-ph/9802302+
    (1998).

  \bibitem{sirlin} A more detailed discussion, which additionally offers 
    a nice historical perspective, can be found in this article:
    Marciano WJ, Sirlin A, \Journal{\PRD}{29}{945}{1984}.

  \bibitem{UA2-Wprod} \uatwo, \Journal{\PLB}{276}{365}{1992}.

  \bibitem{D0-Wprod} \dzero, \Journal{\PRL}{75}{1456}{1995}; 
    \Journal{\PRD}{60}{052003}{1999}; FERMILAB-PUB-99-171-E.

  \bibitem{D0_Ib} \dzero, \Journal{\PRL}{80}{3008}{1998}; 
    \Journal{\PRD}{58}{092003}{1998}.

  \bibitem{Z_mass} The LEP Collaborations \Aleph, \Delphi, \Lt, \Opal\ with
    the LEP Energy Working Group and the \SLD\ Heavy Flavor Group, 
    CERN-EP/2000-016, pp. 5 (2000).

  \bibitem{CDF-95} \cdf, \Journal{\PRL}{75}{11}{1995}; 
    \Journal{\PRD}{52}{4784}{1995}.

  \bibitem{CDF-Wasym_Ia} \cdf, \Journal{\PRL}{74}{850}{1995}.

  \bibitem{CDF-Wasym_I} \cdf,  \Journal{\PRL}{81}{5754}{1998}.

  \bibitem{MRSA} Martin AD, Stirling WJ, and Roberts RG, 
    \Journal{\PRD}{50}{6734}{1994}; \Journal{\PRD}{51}{4756}{1995}.

  \bibitem{CTEQ}  Lai HL etal., \Journal{\PRD}{51}{4763}{1995}.

  \bibitem{LY} Ladinsky GA, Yuan CP, \Journal{\PRD}{50}{4239}{1994}.

  \bibitem{AK} Arnold PB, Reno MH, \Journal{\NPB}{319}{37}{1989}, 
    Erratum \Journal{\ibid}{330}{284}{1990};
    Arnold PB, Kauffman RP, \Journal{\NPB}{349}{381}{1991}.

  \bibitem{CSS} Collins J, Soper D, \Journal{\NPB}{193}{381}{1981}, 
    Erratum \Journal{\ibid}{213}{545}{1983}; 
    Collins J, Soper D, Sterman G, \Journal{\ibid}{250}{199}{1985}.

  \bibitem{AEMG} Altarelli G et al., \Journal{\NPB}{246}{12}{1984}.

  \bibitem{W_rad-BK} 
    Berends FA, Kleiss R, Revol JP, Vialle JP, \Journal{\ZPC}{27}{155}{1985};
    Berends  FA, Kleiss R, \Journal{\ZPC}{27}{365}{1985}.

  \bibitem{W_rad-Baur} Baur U, Berger EL, \Journal{\PRD}{41}{1476}{1990}.

  \bibitem{W_rad2-Baur} Baur U et al., \Journal{\PRD}{56}{140}{1997}; 
    Baur U, Keller S, Sakumoto WK, \Journal {\PRD}{57}{199}{1998}.

  \bibitem{UA2-92} \uatwo, \Journal{\PLB}{276}{354}{1992}.

  \bibitem{HMRSB} Harriman PM, Martin AD, Roberts RG, Stirling WJ, 
    \Journal{\PRD}{42}{798}{1990}; \Journal{\PLB}{243}{421}{1990}.

  \bibitem{CDF-90} \cdf, \Journal{\PRL}{65}{2243}{1990}; 
    \Journal{\PRD}{43}{2070}{1991}.
 
  \bibitem{CDF-99} Lancaster M, 
    FNAL-CONF-99-173-E, to appear in the Proceedings of the XXXIV$^{th}$ 
    Rencontres de Moriond, Les Arcs, France, 1999.

  \bibitem{CDF-Upgrades} \cdf, \Journal{\PRD}{50}{2966}{1994}.

  \bibitem{MRSB} Martin AD, Roberts RG, Stirling WJ, 
    \Journal{\PRD}{37}{1161}{1988}.

  \bibitem{MRSD} Martin AD, Roberts RG, Stirling WJ, 
    \Journal{\PLB}{306}{145}{1993}, 
    Erratum \Journal{\ibid}{309}{492}{1993}.

 \bibitem{MRSR2} Martin AD, Roberts RG, Stirling WJ, 
   \Journal{\PLB}{387}{419}{1996}.

  \bibitem {D0_Ia} \dzero, \Journal{\PRL}{77}{3309}{1996}; 
    \Journal{\PRD}{58}{012002}{1998}.

  \bibitem {D0_EC}\dzero, \Journal{\PRL}{84}{222}{2000}; FERMILAB PUB-99-237-E.

  \bibitem{MRST} Martin AD, Roberts RG, Stirling WJ, Thorne RS,
    \Journal{\EPC}{4}{463}{1998}.

  \bibitem{ybkands} Kunszt Z, Stirling WJ, 1996. In {\it Physics at LEP2},
    volume 1, ed. G Altarelli, T Sj\"{o}strand and F Zwirner, pp. 141-205. 
    Geneva: CERN. 596 pp.

  \bibitem{lep161} \Opal\ Collaboration, \Journal{\PLB}{389}{416}{1996};
    \Delphi\ Collaboration, \Journal{\PLB}{397}{158}{1997};
    \Lt\ Collaboration, \Journal{\PLB}{398}{223}{1997};
    \Aleph\ Collaboration, \Journal{\PLB}{401}{347}{1997}.

  \bibitem{lep172} \Lt\ Collaboration, \Journal{\PLB}{413}{176}{1997};
    \Opal\ Collaboration, \Journal{\EPC}{1}{395}{1998};
    \Aleph\ Collaboration, \Journal{\PLB}{422}{384}{1998};
    \Delphi\ Collaboration, \Journal{\EPC}{2}{581}{1998}.

  \bibitem{lep183} \Opal\ Collaboration, \Journal{\PLB}{453}{138}{1999};
    \Aleph\ Collaboration, \Journal{\PLB}{453}{121}{1999};
    \Lt\ Collaboration, \Journal{\PLB}{454}{386}{1999};
    \Delphi\ Collaboration, \Journal{\PLB}{462}{410}{1999}.

  \bibitem{lep189} Mir L, {\it $W$ Mass from Fully Hadronic Decays at LEP},
     and Chierici R, {\it $W$ Mass from Fully Leptonic and Mixed Decays at
     LEP}, both to appear in the proceedings of EPS'99, Tampere, Finland, 1999.

  \bibitem{ybbandb} Beenaker W, Berends FA, 1996. In {\it Physics at LEP2},
    volume 1, ed. G Altarelli, T Sj\"{o}strand and F Zwirner, pp. 81-140. 
    Geneva: CERN. 596 pp.

  \bibitem{gentle} Bardin D, \etal, {\em Nucl. Phys. Proc. Suppl.} 37B, 
    148 (1994).

  \bibitem{lep_ebm} LEP Energy Working Group, Blondel A, \etal,
    {\it Evaluation of the LEP centre-of-mass energy above the W-pair
    production threshold}, CERN-SL/98-073, accepted by \EPC.

  \bibitem{lepmwth} The LEP Collaborations \Aleph, \Delphi, \Lt, \Opal\ with
    the LEP Energy Working Group and the \SLD\ Heavy Flavor Group, 
    CERN-PPE/97-154, pp. 22-23 (1997).

  \bibitem{durham} Brown N, Stirling WJ, \Journal{\PLB}{252}{657}{1990}.

  \bibitem{lep_rdp} LEP Polarisation Collaboration, 
    \Journal{\PLB}{284}{431}{1992}.

  \bibitem{lep_spec} Torrence E, {\it Determination of the LEP Beam Energy},
    to appear in the proceedings of EPS'99, Tampere, Finland, 1999.

  \bibitem{lep_becr} Ballestrero A, \etal, \Journal{\JPG}{24}{365}{1998}.

  \bibitem{olep1cr} \Opal\ Collaboration, \Journal{\EPC}{11}{217}{1999}.

  \bibitem{o183cr} \Opal\ Collaboration, \Journal{\PLB}{453}{153}{1999}.

  \bibitem{lepmwdr} The LEP Collaborations \Aleph, \Delphi, \Lt, \Opal\ with
    the LEP Energy Working Group and the \SLD\ Heavy Flavor Group, 
    CERN-EP/2000-016, pp. 29-32 (2000).

  \bibitem{LEP-EWK} The LEP Collaborations \Aleph, \Delphi, \Lt, \Opal\ with
    the LEP Energy Working Group and the \SLD\ Heavy Flavor Group, 
    CERN-EP/2000-016 and references therein.

  \bibitem{CCFR} CCFR Collaboration,
    \Journal{\EPC}{1}{509}{1998}.

  \bibitem {NuTeV} McFarland K, Proceedings of the
    XXXIII$^rd$ Rencontres de Moriond, Les Arcs, France 1998. 

  \bibitem {Mtop} \cdf, \Journal{79}{1992}{1997}; 
    \dzero, \Journal{\PRD}{58}{052001}{1998}; 
    \cdf, \Journal{\PRL}{80}{2767}{1998}; 
    \dzero, \Journal{\PRD}{60}{052001}{1999};  
    \cdf, \Journal{\PRL}{82}{271}{1999}; 
    Erratum \Journal{{\ibid}}{82}{2808}{1999}.

  \bibitem{mwmhindirect} The LEP Collaborations \Aleph, \Delphi, \Lt, \Opal\ 
    with the LEP Energy Working Group and the \SLD\ Heavy Flavor Group, 
    CERN-EP/2000-016, pp. 48-49 (2000).

  \bibitem{Degrassi} Degrassi G, Gambino P, Sirlin A, 
    \Journal{\PLB}{394}{188}{1997}; Degrassi G, Gambino P, Passera M, Sirlin A,
    \Journal{\PLB}{418}{209}{1998}.
  
  \bibitem{alpha} Eidelman S, Jegerlehner F, \Journal{\ZPC}{67}{585}{1995}.

  \bibitem{SUSY} Pierce D, Bagger JA, Matchev KT, Zhang RJ, 
    \Journal{\NPB}{491}{3}{1997}.

  \bibitem{TeV2000} {\it Future ElectroWeak Physics at the Fermilab Tevatron:
    Report of the TeV-2000 Study Group}, ed. Amidei D and Brock R, 
    FERMILAB-PUB-96-082 (1996).

  \bibitem{LHCguess} Womersley J, Keller S, \Journal{\EPC}{5}{249}{1998}.

  \bibitem{NLCguess} {\it Physics and Technology of the Next Linear Collider},
    ed. Harris FA, \etal\, SLAC-R-0485 (1996).

  \bibitem{FMCguess} Barger V, Berger MS, Gunion JF, Han T
    \Journal{\PRD}{56}{1714}{1997}.


\end{thebibliography}
\end{document}